\documentclass[letterpaper,twocolumn,10pt]{article}
\usepackage{usenix}
\usepackage{tikz}
\usepackage{amsmath}
\usepackage{lipsum}
\usepackage{booktabs}
\usepackage{lscape}
\usepackage{array,multirow,graphicx,xcolor}
\usepackage{float}
\usepackage{caption}
\usepackage{subcaption}
\usepackage{tcolorbox}
\usepackage{soul}

\begin{document}

\date{}

\title{From Threat Reports to Continuous Threat Intelligence: A Comparison of Attack Technique Extraction Methods from Textual Artifacts}

\author{
{\rm Md Rayhanur Rahman, Laurie Williams}\\
mrahman@ncsu.edu, lawilli3@ncsu.edu \\
North Carolina State University
\and
} 

\maketitle

\begin{abstract}
The cyberthreat landscape is continuously evolving. Hence, continuous monitoring and sharing of threat intelligence have become a priority for organizations. Threat reports, published by cybersecurity vendors, contain detailed descriptions of attack Tactics, Techniques, and Procedures (TTP) written in an unstructured text format. Extracting TTP from these reports aids cybersecurity practitioners and researchers learn and adapt to evolving attacks and in planning threat mitigation. Researchers have proposed TTP extraction methods in the literature, however, not all of these proposed methods are compared to one another or to a baseline. \textit{The goal of this study is to aid cybersecurity researchers and practitioners choose attack technique extraction methods for monitoring and sharing threat intelligence by comparing the underlying methods from the TTP extraction studies in the literature.} In this work, we identify ten existing TTP extraction studies from the literature and implement five methods from the ten studies. We find two methods, based on Term Frequency-Inverse Document Frequency(TFIDF) and Latent Semantic Indexing (LSI), outperform the other three methods with a F1 score of 84\% and 83\%, respectively. We observe the performance of all methods in F1 score drops in the case of increasing the class labels exponentially. We also implement and evaluate an oversampling strategy to mitigate class imbalance issues. Furthermore, oversampling improves the classification performance of TTP extraction. We provide recommendations from our findings for future cybersecurity researchers, such as the construction of a benchmark dataset from a large corpus; and the selection of textual features of TTP. Our work, along with the dataset and implementation source code, can work as a baseline for cybersecurity researchers to test and compare the performance of future TTP extraction methods.
\end{abstract}

\newtcolorbox[auto counter]{mybox}[2][]{%
title=Example~\thetcbcounter: #2, #1}

\section{Introduction}

Information technology (IT) systems have been gaining continuous attention from threat actors with financial motives~\cite{hackernews-financial-backup} and organized backing (i.e., state sponsored~\cite{reuters}). For example, in 2021, Sonatype reported that software supply chain attacks increased by 650\% in 2020 from the previous year~\cite{sonatype}. Moreover, a cyberattack on the Colonial pipeline~\cite{colonial-pipeline}, JBS~\cite{jbs-attack} and Ireland health services~\cite{irish-health-attack} show that threat actors can destabilize millions of people's lives by fuel price surge, food supply shortage, and disruption in healthcare services. Thwarting cyberattacks has become more complicated as the threat landscape evolves rapidly. Hence, continuous monitoring and sharing of threat intelligence has become a priority, as emphasized in Section 2(iv) of the US Executive Order 14028: Improving the Nation's cybersecurity: ``service providers share cyber threat and incident information with agencies, doing so, where possible, in industry-recognized formats for incident response and remediation.``~\cite{useo}.

Threat reports, published by cybersecurity vendors and researchers, contain detailed descriptions on how malicious actors utilize specific tactics, relevant techniques, and describe procedures for performing the attack - known as Tactics, Techniques, Procedures (TTP) (see Section~\ref{ttps})~\cite{tounsi2018survey, ti-dns-stuff, attack}) - to launch cyberattacks. Consider a threat report from FireEye describing the attack procedures of the Solarwinds supply chain attack~\cite{fireeye} in Example 1, where we show the attackers' actions in bold text. One of the observed (mentioned in the report) TTP is \textit{T1518.001: Security Software Discovery} which allows an attacker to bypass the security defense by discovering security software running in the system~\cite{ttp-software-discovery}. The rise in cyberattack incidents with evolving attack techniques results in a growing number and volume of threat reports. Extracting the TTP from threat reports can help cybersecurity practitioners and researchers with cyberattack characterization, detection, and mitigation~\cite{husari2017ttpdrill} from the past knowledge of cyberattacks. Analyzing TTP also helps cybersecurity practitioners in continuous monitoring and sharing of threat intelligence. For example, organizations can learn how to adapt to the evolution of cyberattacks. Cybersecurity red and blue teams also benefit in threat hunting by threat intelligence sharing~\cite{tounsi2018survey}, attack profiling~\cite{noor2019machine}, and forecasting~\cite{sabottke2015vulnerability}.

Threat reports contain a large amount of text and manually extracting TTP is error-prone and inefficient~\cite{husari2017ttpdrill}. Cybersecurity researchers have proposed automated extraction of TTP from threat reports (e.g.~\cite{rahman2021attackers, husari2017ttpdrill, husari2018using, niakanlahiji2018natural, ayoade2018automated, noor2019machine}). Moreover, the MITRE~\cite{mitre-org} organization uses an open-source tool~\cite{mitre-tram} for finding TTP from threat reports. These TTP extraction work use natural language processing (NLP) along with supervised and unsupervised machine learning (ML) techniques to classify texts to the corresponding TTPs. However, no comparison among this existing work has been conducted, and the research has not involved an established ground truth dataset~\cite{rahman2021attackers}, highlighting the need for a comparison of underlying methods of existing TTP extraction work. A comparative study among these methods would provide cybersecurity researchers and practitioners a baseline for choosing the best method for TTP extraction, finding room for improvement. 



Rahman et al. systematically surveyed the literature and obtained ten TTP extraction studies  ~\cite{rahman2021attackers}. None of these studies compared their work with a common baseline and only two of these studies~\cite{ayoade2018automated, husari2018using} compared their results with one other. In our work, we first select five studies~\cite{noor2019machine, niakanlahiji2018natural, husari2017ttpdrill, ayoade2018automated,li2019extraction} from the ten based on inclusion criteria (Section~\ref{filter}) and implement the underlying methods of the five selected studies. We then compare the performance of classifying text (i.e., attack procedure description) to the corresponding attack techniques. Moreover, as the number of attack techniques are growing due to the evolution of attack techniques, we also investigate (i) how the methods perform given that the dataset has class imbalance problems (existence of majority and minority classes); and (ii) how the methods perform when we increase the classification labels (labels are the name of techniques that would be classified from attack procedure descriptions). 

\textit{The goal of this study is to aid cybersecurity researchers and practitioners choose attack technique extraction methods for monitoring and sharing of threat intelligence by comparing underlying method from the TTP extraction studies in the literature.}  We investigate these following research questions (RQs):

\begin{description}
    \item[RQ1: Classification performance] How do the TTP extraction methods perform in classifying textual descriptions of attack procedures to attack techniques across different classifiers?
    \item[RQ2: Effect of class imbalance mitigation] What is the effect on the performance of the compared TTP extraction methods when oversampling is applied to mitigate class imbalance?
    \item[RQ3: Effect of increase in class labels] How do the TTP extraction methods perform when the number of class labels is increased exponentially?
\end{description}

We implement the underlying methods of these five studies: ~\cite{noor2019machine, niakanlahiji2018natural, husari2017ttpdrill, ayoade2018automated,li2019extraction}. We construct a pipeline for comparing the methods on the same machine learning workflow.  We run the comparison utilizing a dataset constructed from the MITRE ATT\&CK framework~\cite{attack}. We also run the methods on oversampled data to investigate how the effect of class imbalance can be mitigated. Finally, we use six different multiclass classification settings ($n = 2,4,8,16,32,64$ where $n$ denotes the number of class labels) to investigate how the methods perform in classifying a large number of available TTP. We list our contributions below.

\begin{itemize}
    \item A comparative study of the five TTP extraction methods from the literature. This article, to the best of our knowledge, is the first study to conduct direct comparisons of the TTP extraction methods.
    \item A sensitivity analysis on the effect of using oversampling and multiclass classification on the compared method. Our work investigates these two important aspects of classification as the number of techniques is more than hundred and the technique enumeration is being updated gradually resulting in majority and minority classes.   
    \item A pipeline for conducting the comparison settings which ensure the methods are executed in the same machine learning workflow. We also make our dataset and implementation source code available at~\cite{githubrepo} for future researchers. The pipeline along with the dataset and implementation sources serve as a baseline for cybersecurity researchers to test and compare the performance of the future TTP extraction method.
    \item We provide recommendations on how the methods can be improved for better extraction performance. 
\end{itemize}

\begin{mybox}{Excerpt from a threat report on Solarwinds attack showing attackers' actions in bold texts}
After an initial dormant period of up to two weeks, it \textbf{retrieves} and \textbf{executes} commands, called “Jobs”, that include the ability to \textbf{transfer} files, \textbf{execute} files, \textbf{profile} the system, \textbf{reboot} the machine, and \textbf{disable} system services. The malware \textbf{masquerades} its network traffic as the Orion Improvement Program (OIP) protocol and \textbf{stores} reconnaissance results within legitimate plugin configuration files allowing it to blend in with legitimate SolarWinds activity. The backdoor uses multiple obfuscated blocklists to \textbf{identify} forensic and anti-virus tools running as processes, services, and drivers.
\tcblower
Source: FireEye~\cite{fireeye}\\
\end{mybox}

The rest of the article is organized as follows. In Section~\ref{background}, we discuss a few key concepts relevant to this study. In Section~\ref{selection} and~\ref{work_overview}, we discuss our process to identify the selected studies for comparison. In Section~\ref{design}, we discuss our methodology for designing and running the experiment. In Section~\ref{finding} and ~\ref{discussion}, we report and discuss our observations from the experiment. In Section~\ref{limitation} and ~\ref{future}, we identify several limitations to our work followed by highlighting potential future research paths. In Section~\ref{related-work}, we discuss related work in the literature followed by concluding the article in Section~\ref{conclusion}. We report a few supplementary information in the Appendix.

\section{Key Concepts} 
\label{background}
In this section, we discuss several key concepts relevant in the context of our study. 

\subsection{Threat Intelligence:} Threat intelligence - also known as Cyberthreat intelligence (CTI) - is defined as `evidence-based knowledge, including context, mechanisms, indicators, implications, and actionable advice about an existing or emerging menace or hazard to assets that can be used to inform decisions regarding the subject’s response to that menace or hazard`~\cite{mcmillan2013definition}. Threat intelligence can be used to forecast, prevent and defend attacks.

\subsection{Tactics, techniques and procedures (TTP):} 
\label{ttps}
Tactics are high level goals of an attacker, whereas techniques are lower level descriptions of the execution of the attack in the context of a given tactic~\cite{attack, tounsi2018survey}. Procedures are the lowest level step by step execution of an attack being performed. TTP can be used to profile or analyze the lifecycle of an attack on a targeted system. For example, privilege escalation is a tactic for gaining elevated permission on a system. One technique for privilege escalation can be access token manipulation~\cite{attack}. An attacker can gain elevated privilege in a system by tampering the access token to bypass the access control mechanism. An example procedure is an attacker manipulating an access token by using Metasploit's named-pipe impersonation~\cite{attack}. 

\subsection{ATT\&CK:} The MITRE~\cite{mitre-org} organization developed ATT\&CK~\cite{attack}, a framework derived from real world observations of adversarial TTPs deployed by attack groups. ATT\&CK contains an enumeration of high level attack stages known as tactics. Each tactic has an enumeration of corresponding techniques, and each technique has associated procedure description(s). Procedures are written in unstructured text and describe how a particular technique has been used by the attacker to gain an objective of the corresponding tactic to launch a cyberattack. ATT\&CK was first introduced in 2013 to model the lifecycle and common TTP utilized by threat actors in launching APT (advanced persistent threat) attacks. In our research, we utilized Version 9 of the ATT\&CK framework which consists of 14 Tactics, 170 Techniques, and 8,104 procedures. 




\section{Selection of TTP extraction methods}
\label{selection}
In this section, we discuss the methodology for selecting and comparing the TTP extraction methods in five studies ~\cite{noor2019machine, niakanlahiji2018natural, husari2017ttpdrill, ayoade2018automated,li2019extraction} found in literature.

\subsection{Finding TTP extraction work from the literature:} Rahman et al.~\cite{rahman2021attackers} systematically collected automated threat intelligence extraction-related studies from scholarly databases and found 64 relevant studies. From these, the first author of this paper identified ten studies that extracted TTP from the text automatically using NLP and ML techniques. We select these ten work as potential candidates for our comparison study. We refer to these ten works as the \textit{candidate set}. In the Appendix, Table~\ref{tab:candidate-studies}, we list the bibliographic information of the candidate set. 

\begin{table*}[htb]
    \centering
    \footnotesize
    \begin{tabular}{lp{4cm}p{4cm}rp{4cm}}
        \toprule
        \textbf{Id} & \textbf{Dataset type} & \textbf{Dataset source} & \textbf{\# threat reports} & \textbf{NLP/ML techniques and features} \\ \midrule
        $S_1$* & Data breach incident reports & Github APTnotes~\cite{aptnotes} and custom search engine~\cite{cse} & 327 & Latent Semantic Indexing(LSI) \\
        $S_2$* & APT attack reports & Github APTnotes & 445 & Dependency parsing, TFIDF of independent noun phrases \\
        $S_3$ & APT attack reports & Github APTnotes & 50 & Named entity recognition(NER) \\
        $S_4$* & APT attack reports & Github APTnotes & 18,257 & TFIDF \\
        $S_5$ & Malware report & Github APTnotes, MicrosoftS/Adobe Security Bulletins, National Vulnerability Database description & 474 & NER, Cybersecurity ontology \\
        $S_6$* & APT attack reports & Attack technique dataset (Source not reported) & 200 & LSI \\
        $S_7$ & Computer security literature and Android developer documentation & IEEE S\&P, CCS, USENIX articles, Android API~\cite{android}  & 1,068 & Dependency parsing \\
        $S_8$ & - & - & 18 & NER, Dependency parsing, Basilisk \\
        $S_9$* & Malware report & Symantec threat reports& 17,000 & Dependency parsing, BM25 \\
        $S_{10}$ & Malware report & Symantec threat reports& 2,200 & Dependency parsing, BM25 \\
        \bottomrule
        \multicolumn{5}{c}{Id with (*) symbol denotes that the study is selected for comparison}
    \end{tabular}
    \caption{Datasets and methods used in candidate set}
    \label{tab:candidate-set-details}
\end{table*}

\subsection{Inclusion criteria for TTP extraction work:} 
\label{filter}
A comprehensive comparison of TTP extraction methods is not a straightforward task. One difficulty in setting up the study is to find a labelled and universally agreed upon dataset. Moreover, constructing such a dataset is inherently challenging as the set of TTP is subject to change with evolution of the manner of attack. Another challenge is to determine whether the extraction should be performed on the sentence level or paragraph level. Finally, in the candidate set, TTP extraction methods were designed targeting different use cases, such as transforming the extracted TTP to structured threat intelligence formats~\cite{husari2017ttpdrill} or building a knowledge graph~\cite{piplai2020creating}. Hence, not every study in the candidate set is able to extract all known TTPs. Hence, we define the following inclusion criteria: 

\begin{enumerate}
    \item All methods selected for the comparison can work on the same textual artifacts
    \item Besides labelling the text to corresponding technique, no other manual labelling is required for comparison
    \item All methods can be compared using the same set of technique names which will be used as labels for classification tasks.
\end{enumerate}

\subsection{Filtering the TTP extraction work for comparison:} In Table~\ref{tab:candidate-set-details}, we report the dataset type, dataset source, and relevant NLP/ML techniques used for our candidate set. Next, we report how we filter the candidate set. 

\begin{itemize}
    \item We drop $S_3$, $S_5$, and $S_8$ because Named Entity Recognition (NER) labelling of words from the text is required (violates filtering criteria [2]).
    \item We drop $S_7$ because this work (a) uses Android development documentation (violates filtering criteria [1]), and (b) extracts the features for Android-specific malware only (violates filtering criteria [3]).
    \item We drop $S_{10}$ because the work requires additional manual work on identifying relevant verbs and objects from Wikipedia articles on computing and cybersecurity related concepts (violates filtering criteria [2]).
\end{itemize}

Finally, we keep the remaining work for our comparison study: $S_1$, $S_2$, $S_4$, $S_6$, and $S_9$. $S_1$ and $S_6$ utilized Latent Semantic Indexing (LSI)~\cite{landauer1998introduction}; $S_2$ and $S_4$ utilized Term frequency - inverse document frequency (TFIDF); and $S_9$  utilized dependency parsing and BM25. 

\section{Overview of the selected studies for comparison}
\label{work_overview}
We report a brief overview of the studies selected for comparison followed by observed similarities and dissimilarities.

\begin{description}
  \item[$S_1$] The authors used the data breach incident reports produced by cybersecurity vendors and then searched high level attack patterns from those reports. The authors used the ATT\&CK framework for the common vocabulary of attack pattern names. They used LSI for searching the attack pattern names from the texts. Finally, they correlated these searched attack patterns with responsible APT actor groups. 
  
  \item[$S_2$] The authors used APT attack related articles as dataset and MITRE ATT\&CK framework for the common vocabulary of TTP. Then they extracted independent noun phrases from the corpus that appear in the corpus at least once without being part of a larger noun phrase. Then they computed TFIDF vectors of these noun phrases. Finally, using these vectors, they retrieved the most relevant set of articles associated with specific TTP keywords such as data breach, privilege escalation.
  
  \item[$S_4$] The authors used APT attack-related articles and Symantec threat reports as dataset and MITRE ATT\&CK framework for the common vocabulary of TTP. They computed TFIDF vectors of the articles and then applied three bias correction techniques named kernel mean matching~\cite{gretton2009covariate}, Kullback-Liebler importance estimation procedure~\cite{sugiyama2008direct}, and relative density ratio estimation. Finally, they used SVM classifier on bias corrected data. 
  
  \item[$S_6$] The authors used advanced persistent threat (APT) attack related online articles as dataset and MITRE ATT\&CK framework for the common vocabulary of TTP. They first computed the TFIDF vectors of the description of TTP. Then they applied LSI on articles for retrieving a set of topics. After that, for each article, the authors computed the cosine similarity score between TFIDF vectors of each TTP and the retrieved topics. Then they used these computed similarity scores as features. Finally, the authors used two multi-label classification techniques named Binary Relevance and Label Powerset~\cite{read2011classifier, spolaor2013comparison}. 
  
  \item[$S_9$] The authors used Symantec threat reports as dataset and MITRE ATT\&CK framework for the common vocabulary of TTP. First, they created an ontology of threat actions from the description of ATT\&CK. Then they extracted threat actions as (subject, verb, object) tuples from each sentence in the corpus. Finally, they computed BM25 score for threat actions against their created ontology and mapped the extracted threat actions to the corresponding entities in their ontology and converted the reports to the Structured Threat Intelligence Exchange (STIX~\cite{stix}) format. 
\end{description}

From the above description, we see the following similarities among the articles: 

\begin{itemize}
    \item Threat reports are used as dataset,
    \item The MITRE ATT\&CK framework is used as the common vocabulary of TTP
    \item NLP techniques (such as TFIDF, LSI, BM25) are used for feature extraction or computation from text.
    \item The extracted or computed textual features are fed to machine learners.
\end{itemize}

However, we also observed the following dissimilarities among the studies as well: 

\begin{itemize}
    \item The purpose of the TTP extraction is different. For example, $S_1$ correlated extracted TTP with APT actor groups while $S_9$ constructed STIX threat intelligence format from unstructured threat reports
    \item Not all studies used classification techniques, such as $S_1$, $S_2$
    \item Bias correction on dataset is only applied in $S_4$
    \item Only one study ($S_6$) modelled their approach as multi label classification problem
\end{itemize}

\section{Comparison Study}
\label{design}
In this section, we provide our research steps for the comparison study. We report how we design and utilize a five-step TTP extraction pipeline (see Section~\ref{pipeline}) for running the comparison study.  We construct our dataset from MITRE ATT\&CK framework. Next, we apply pre-processing to the corpus. After that, we implement the underlying method of the five selected studies. Finally, we feed the extracted/computed features to classifiers with oversampling (see Section~\ref{oversampling}) and multiclass classification (see Section~\ref{multiclass}) settings.  

\subsection{Comparison scope} \label{scope}
In Section~\ref{work_overview}, we report similarities and dissimilarities among the articles we choose for comparison and as a result, we define the following scope of the comparison.

\begin{itemize}
    \item We mention in Section~\ref{background} that ATT\&CK contains the mapping between (a) tactics and techniques, and (b) techniques and textual descriptions of procedures. Classifying the procedure description to the corresponding technique would find the corresponding tactics as well as ATT\&CK provides the tactics-technique mapping as part of the framework. Hence, for TTP extraction, we choose to classify techniques from a given text (which are procedure descriptions), which will also give us the associated tactics from the tactic-technique mapping. 
    \item We assume that a piece of text (i.e., sentence(s)) is related to one technique. Hence, we will use classifiers to classify a piece of text to its corresponding technique. 
    \item All methods will be compared using the same dataset, class labelling, same set of classifiers
    \item All the methods will be compared in the same machine learning pipeline (See Section~\ref{pipeline})
\end{itemize}

As we will use the same dataset, labels, and classifiers, hence, we will only compare the NLP methods for extracting textual features that we will feed towards classifiers.
    \begin{enumerate}
        \item For $S_1$, we will use LSI vectors as features. We will refer this method as M:LSI.
        \item For $S_2$, we will use TFIDF of unique noun phrases (See Section~\ref{work_overview}) as features. We will refer this method as M:TFIDF-NP.
        \item For $S_4$, we will use TFIDF vectors of the corpus as features. We will refer this method as M:TFIDF.
        \item For $S_6$, we will use the cosine similarity score as features. The score is computed from TFIDF vector of each technique description and vectors of topics generated by LSI. We will refer this method as M:LSI-Co.
        \item For $S_9$, we will use BM25 score of (subject, verb, object) tuples of TTP description and corpus texts as features. We will refer this method as M:BM25.
    \end{enumerate}

\subsection{Metrics for performance measurement}
We use the following metrics for measuring the performance of TTP extraction methods we are comparing.

\begin{description}
  \item[Precision] the ratio of true positives and the sum of true positives and false positives indicating the relevance of the performed classification. The higher the precision score is, the better the classifier is in finding relevant examples from a given class.
  \item[Recall] the ratio of true positives and the sum of true positives and false negatives indicating the completeness of the performed classification. The higher the recall score is, the better the classifier is in finding all examples from a given class.
  \item[F1 score] The harmonic mean of precision and recall indicating how good the classification task in terms of both precision and recall. The higher the F1 score is, the better the classifier is in both classifying relevant examples to correct classes and classifying all examples correctly belonging to the same class.
  \item[AUC score] the area under the curve of Receiver Operating Characteristics indicating the ability of a classifier  to separate the true positive and true negative examples. Higher AUC score indicates better performance in classification. AUC score equalling $0.5$ denotes that the classifier is as good as a random guess only, less than $0.5$ means worse than a random guess. AUC closer to $1$ is desirable and a higher AUC score indicates the classifier is able to choose a randomly selected true positive example with higher confidence than choosing a true negative example. 
\end{description}


\begin{figure*}
    \centering
    \includegraphics[width=\textwidth]{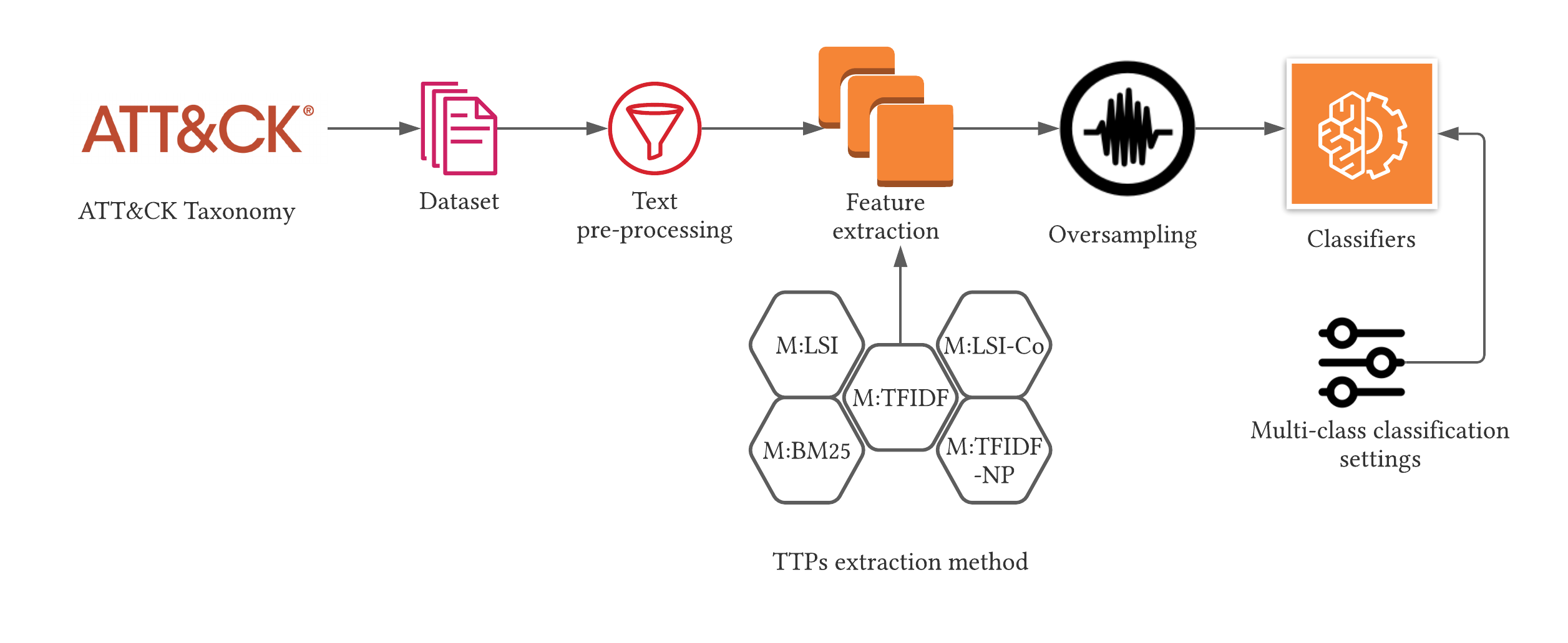}
    \caption{Instantiated pipeline for our comparison experiment}
    \label{fig:pipeline}
\end{figure*}

\subsection{TTP extraction pipeline for comparison} \label{pipeline}
We construct a TTP extraction pipeline for comparing the methods through five steps, as defined below. 

\textbf{Step 1: dataset collection} In this step, we collect the dataset (Section~\ref{dataset}) for the comparison experiment. The dataset contains (a) the textual description of attack procedures;  and (b) ground truth for the classification task which are the corresponding attack techniques used in the description.

\textbf{Step 2: text pre-processing} In this step, we apply the following on the corpus: (a) pre-processed for filtering the punctuation marks and whitespaces; (b) tokenization; (c) stop word removal; (d) stemming; and (e) lemmatization.

\textbf{Step 3: feature extraction} The underlying methods of the selected TTP extraction work (see Section~\ref{work_overview}) varies from one another in the context of NLP techniques used and the choice of the classifiers. Hence, in this step, we implement the methods we discuss in Section~\ref{work_overview} to extract the textual features that we would provide to classifiers.

\textbf{Step 4: oversampling (optional)} In this step, we apply oversampling to our dataset to mitigate the introduced bias by class imbalance in the dataset. This step is optional and the methods would be compared both with this step and without this step. 

\textbf{Step 5: classification} In this step, we train the classifiers with a portion from our dataset and then test the classification performance with the rest of the dataset that we do not use for training purposes. 

We instantiate these five steps in the following five subsections.  Figure~\ref{fig:pipeline} shows the instantiated pipeline.

\begin{table*}[htb]
    \centering
    \footnotesize
    \begin{tabular}{lp{8cm}lp{4cm}}
        \toprule
        \textbf{Procedure Id} & \textbf{Procedure description} & \textbf{Technique Id} & \textbf{Technique name} \\ \midrule
        G0016 & APT29 has used encoded PowerShell scripts uploaded to CozyCar installations to download and install SeaDuke. APT29 also used PowerShell to create new tasks on remote machines, identify configuration settings, evade defenses, exfiltrate data, and to execute other commands. & T1059 & Command and Scripting Interpreter \\ \midrule
        
        S0045 & Most of the strings in ADVSTORESHELL are encrypted with an XOR-based algorithm; some strings are also encrypted with 3DES and reversed. API function names are also reversed, presumably to avoid detection in memory & T1027 & Obfuscated Files or Information \\ \midrule
        
        S0154 & Cobalt Strike can conduct peer-to-peer communication over Windows named pipes encapsulated in the SMB protocol. All protocols use their standard assigned ports. & T1071 & Application Layer Protocol \\ \midrule
        
        G007 & APT28 has downloaded additional files, including by using a first-stage downloader to contact the C2 server to obtain the second-stage implant & T1105 & Ingress Tool Transfer \\ \midrule
        
        S0449 & Maze has used the "Wow64RevertWow64FsRedirection" function following attempts to delete the shadow volumes, in order to leave the system in the same state as it was prior to redirection & T1070 & Indicator Removal On Host \\
        \bottomrule
    \end{tabular}
    \caption{Example of attack procedure description and corresponding technique in our dataset}
    \label{tab:dataset-examples}
\end{table*}


\subsection{Step 1:  dataset collection}\label{dataset}
We construct the dataset from the textual description of attacks procedures mentioned in the MITRE ATT\&CK framework. In the MITRE ATT\&CK framework, each of the tactics has a one-to-many mapping with techniques and each of these techniques has a one-to-many mapping with textual descriptions of procedures taken from real-world cybersecurity incidents described in threat reports. We choose MITRE ATT\&CK framework for constructing the dataset for the following reasons:
\begin{itemize}
    \item Although there is an abundance of threat reports in the internet, threat reports have to be manually filtered for relevance and manual labelled to TTP for each sentence of those threat reports. On the other hand, in the ATT\&CK framework, the mapping between techniques and descriptions of procedures are already present and these mappings have been performed by cybersecurity professionals and researchers. The procedure-technique mapping done by these professionals are our ground truth. 
    \item All studies we selected for comparison use the ATT\&CK framework for the common vocabulary of attack technique names. 
    \item The dataset contains textual descriptions of the attack procedures listed in the ATT\&CK framework. The textual description of attack procedures consists of a few sentences and has the mapping to the associated technique name. 
    \item The ATT\&CK framework is regularly updated and maintained with the evolution of attacks, hence, we get the latest set of TTP with the latest version of the ATT\&CK framework. 
\end{itemize}

We use Version 9 of the ATT\&CK framework ~\cite{attack-dataset}. The dataset contains 8,104 attack procedure descriptions and corresponding techniques used in a real world cyber attack. The dataset contains 170 techniques. However, some techniques have more than hundreds of procedure descriptions while other techniques have only one. In our study, we use techniques that have at least 30 procedure descriptions, resulting in 7,061 procedure descriptions and 64 techniques. In Table~\ref{tab:dataset-examples}, we show a few examples of techniques and textual description of procedures. For each of the textual descriptions, the mapped technique name will be used as the label for the classification task. 


\subsection{Step 2:  text pre-processing}
We first remove the urls and citations. Then, we use the \verb|gensim.parsing.preprocessing| from \verb|gensim| Python library for removing punctuation, whitespaces, stop-words and performing tokenization, stemming and lemmatization. 

\subsection{Step 3: feature extraction} 
\begin{description}
    \item[M:TFIDF] We compute the TFIDF vectors of the corpus. Then, we normalize the TFIDF vectors to unit length. Finally, we feed these normalized TFIDF vectors to classifiers. We use \verb|TfidfVectorizer| from the \verb|scikit-learn| Python package. 

    \item[M:LSI] We first compute the TFIDF vectors of the corpus. Then we normalize the TFIDF vectors to unit length. Then, we apply LSI to apply dimensionality reduction of the computed vectors. Finally we feed these normalized TFIDF vectors to classifiers. We use \verb|tfidfmodel| and \verb|lsimodel| from \verb|gensim| Python packages.

    \item[M:LSI-Co] We apply pre-processing on the corpus as well as textual description of the techniques mentioned in the ATT\&CK framework. Then, we compute the TFIDF vector for each of the technique description in the ATT\&CK framework followed by normalization. After that, we apply LSI to each textual description in the corpus. We then compute the cosine similarity between TFIDF vectors (of each of the techniques) and topic vectors (of each of the textual descriptions from the corpus). Finally, we use these cosine similarities as features that we will feed towards classifiers. We use \verb|tfidfmodel|, \verb|lsimodel| from \verb|gensim|, and \verb|scipy| Python packages.

    \item[M:TFIDF-NP] We use parts of speech tagging to determine the noun phrases in the corpus. Then we construct a list of all noun phrases identified in the corpus. After that, we identify the noun phrases from the list which have been found in the corpus at least once without being part of a larger noun phrase. Then we compute the TFIDF vectors of these independent noun phrases for each procedure description. Then we normalize the TFIDF vectors to unit length. Finally, we feed these normalized TFIDF vectors to classifiers. We use \verb|TfidfVectorizer| from \verb|scikit-learn| and \verb|spacy| Python package. 

    \item[M:BM25] We first apply pre-processing to the corpus and textual description of the techniques from the ATT\&CK framework. We then apply dependency parsing to both the corpus and textual description of the techniques from the ATT\&CK framework. After that, we then extract the \verb|(subject, verb, object)| (SVO) tuples from both the corpus and textual description of techniques from the ATT\&CK framework. We randomly select 100 procedure descriptions and the first two authors manually verified the tuple extraction performance. The first author finds 74\% of the tuples extracted and the second author finds 88\% of the tuples extracted by our implementation. Next, we construct Bag-of-words representation of the extracted \verb|(subject, verb, object)| tuples from both the corpus and textual description of techniques from the ATT\&CK framework. After that, we compute the similarity score using BM25 ranking method for the extracted \verb|(subject, verb, object)| tuples for each corpus and \verb|(subject, verb, object)| tuples for each technique from ATT\&CK framework. Finally, we use these similarity scores as features that we would feed to the classifiers. We use \verb|spacy|, \verb|BM25Okapi|.

\end{description}

\subsection{Step 4: oversampling (optional)} \label{oversampling}
As we mention in Section~\ref{dataset}, in our dataset, the techniques do not have the same amount of procedure descriptions. Some techniques have more than hundreds and some of the techniques have only 30. As a result, there are majority and minority classes in the dataset leading to class imbalance in the dataset. We use oversampling to mitigate the class imbalance problem in our dataset, and we apply a technique called SMOTE which stands for Synthetic Minority Oversampling Technique~\cite{chawla2002smote}. SMOTE works by choosing a random example from a minority class and then selecting the nearest $k$ neighbors of that minority class to generate more synthetic examples of that minority class. We ran all methods with and without oversampling technique to observe their performance with and without handling the class imbalance issue. As SMOTE can only apply oversampling to numeric features, we applied SMOTE on the computed features in each method, such as TFIDF vectors, similarity score. We use \verb|SMOTE| Python package to implement the oversampling. 

\begin{table}[]
    \centering
    \footnotesize
    \begin{tabular}{ccc}
    \toprule
         \multirow{2}{*}{\textbf{\# Classes}} & \multicolumn{2}{c}{\textbf{\# Descriptions}} \\ \cline{2-3}
         {} & \textbf{Before oversampling} & \textbf{After oversampling} \\ \midrule
         $n = 2$ & 890 & 1,064 \\
         $n = 4$ & 1,492 & 2,128 \\
         $n = 8$ & 2,448 & 4,256 \\
         $n = 16$ & 3,688 & 8,512 \\
         $n = 32$ & 5,390 & 17,024 \\
         $n = 64$ & 7,061 & 34,048 \\
     \bottomrule
    \end{tabular}
    \caption{Procedure description count for each multiclass classification setting before and after oversampling}
    \label{tab:dataset-size-oversampled}
\end{table}

\subsection{Step 5:  classification}
We use six classifiers named Naive Bayes(NB), Support Vector Machine(SVM), Neural Network(NN), K-nearest Neighbor(KNN), Decision Tree(DT) and Random Forest(RF) classifiers. Collectively, these six classifiers are used by the authors of the study selected for the comparison. We use \verb|scikit-learn| Python package for these classifiers and we use \verb|GaussianNB|, \verb|SVC|, \verb|KNeighborsClassifier|, \verb|MLPClassifier|, \verb|DecisionTreeClassifier|, and \verb|RandomForestClassifier| for NB, SVM, KNN, NN, DT, and RF classifiers respectively. 

\subsection{Multi-class classification settings} \label{multiclass}
We mention in Section~\ref{dataset} that our dataset contains 64 technique names. Hence, in a corpus, a piece of text can be classified to one of the maximum 64 possible technique names. Thus, in this experiment, the TTP extraction can be considered a multi-class classification problem. We run each of the methods with the classifiers in these two following cases: (a) where a piece of text can be classified to one from two possible technique names, and (b) where a piece of text can be classified to one from more than two possible technique names to observe how the method performs in both cases. For case (a), we sort the technique names by the count of the corresponding procedure descriptions in the dataset, and then we select the top two techniques and their corresponding descriptions. For case (b), we we sort the technique names by the count of the corresponding procedure descriptions in the dataset, and then we select the top $n$ (here, $n$ is the number of possible classification labels) techniques and their corresponding descriptions. We run all methods in the following six different cases ($n = 2,4,8,16,32,64$ where $n$ denotes the number of class labels) which we will refer as multiclass classification settings settings. We report the procedure description count for each of the multiclass classification settings before and after oversampling in Table~\ref{tab:dataset-size-oversampled}.


\subsection{Cross-validation}
We apply K-fold cross-validation technique to split our dataset into different sets of training and testing set. We choose K=5 and for each fold, we use $80\%$ of the dataset as training set and rest of the dataset ($20\%$) as testing set. 

\subsection{Hyperparameters}
We use the default hyperparameter settings reported in the \verb|scikit-learn| package for the classifiers. For \verb|SMOTE|, we choose $k=6$ for generating synthetic examples from the nearest neighbors. Finally, before executing any method, we set the \verb|np.random.RandomState| property of the Python environment to $0$ to ensure the replicability of our results from execution. For determining the number of topics for LSI model, we use the \verb|coherencemodel.get_coherence()| method from \verb|lsimodel| to get the coherence value for \verb|num_topic| from 1 to 500. We observe that the coherence value increases monotonically. Hence, for LSI models, we use the \verb|num_topic = 500|. 

\section{Results}
\label{finding}
In this section, we report the findings for each of the RQs. 

\subsection{RQ1: Classification performance}
We report the precision, recall, F1 score, and AUC scores for all implemented methods paired with each of the six classifiers in Table~\ref{tab:rq1}. Each corresponding cell in the table reports the score in $A-B(C)$ format where $A$ is the minimum observed score, $B$ is the maximum observed score, and $C$ is the arithmetic average of a method run with a classifier in all multiclass classification settings (see Section~\ref{multiclass}): $n = 2, 4, 8, 16, 32, 64$ where $n$ denotes the number of class labels in the dataset. For example, the top left cell containing $63-88(76)$ denotes the precision score of M:TFIDF method paired with KNN classifier. The minimum and maximum precision scores observed are $63$ and $88$, respectively, from all of six possible multi-class classification settings ($n = 2, 4, 8, 16, 32, 64$), and the average precision score observed is $76$. We bold the cell which shows the maximum average score for each method paired with six classifiers. We report the performance score of all methods across six classifiers, all oversampling settings, and all multiclass classification settings at Table~\ref{tab:whole-report} in Appendix.

\renewcommand{\tabcolsep}{2pt}
\begin{table}[t]
\centering
\footnotesize
\begin{tabular}{c|c|cccc}
\toprule
\textbf{M}     & \textbf{C}          & \textbf{P}         & \textbf{R}         & \textbf{F}         & \textbf{A}         \\ \midrule
\parbox[t]{2mm}{\multirow{1}{*}{\rotatebox[origin=c]{90}{M:TFIDF}}}    
           & KNN        & 63-88(76) & 55-84(71) & 56-85(71) & 88-93(91) \\ 
           & NB         & 28-71(41) & 25-71(40) & 25-71(40) & 62-71(64) \\ 
           & SVM        & \textbf{77-92(87)} & 65-90(81) & 68-90(83) & \textbf{98-99(99)} \\ 
           & DT         & 58-89(77) & 56-84(74) & 56-84(74) & 80-88(85) \\ 
           & RF         & 72-92(85) & 66-89(82) & 67-89(82) & 98-99(98) \\ 
           & NN         & 74-91(86) & \textbf{71-90(84)} & \textbf{71-90(84)} & 97-99(98) \\ \midrule
\parbox[t]{2mm}{\multirow{1}{*}{\rotatebox[origin=c]{90}{M:TFIDF-NP}}}
           & KNN        & 39-77(59) & 33-74(55) & 33-73(55) & 79-86(84) \\ 
           & NB         & 31-83(54) & 35-76(53) & 30-77(51) & 76-88(82) \\
           & SVM        & 42-80(63) & 28-78(55) & 30-77(56) & 88-89(89) \\ 
           & DT         & 48-81(68) & 46-78(66) & 45-77(66) & 77-86(83) \\ 
           & RF         & \textbf{58-84(75)} & \textbf{53-82(73)} & \textbf{53-81(72)} & \textbf{93-96(95)} \\ 
           & NN         & 57-84(73) & 52-82(71) & 53-81(71) & 92-94(93) \\ \midrule
\parbox[t]{2mm}{\multirow{1}{*}{\rotatebox[origin=c]{90}{M:LSI}}}
           & KNN        & 61-84(71) & 51-81(64) & 53-81(65) & 86-89(87) \\ 
           & NB         & 62-72(67) & 55-69(61) & 40-69(59) & 62-94(86) \\ 
           & SVM        & \textbf{75-92(87)} & \textbf{67-90(83)} & \textbf{69-90(83)} & \textbf{98-98(99)} \\ 
           & DT         & 35-87(66) & 35-86(66) & 35-86(66) & 71-86(81) \\ 
           & RF         & 67-90(83) & 55-86(77) & 57-86(78) & 95-98(97) \\ 
           & NN         & 72-86(80) & 70-85(78) & 70-84(77) & 82-98(94) \\ \midrule
\parbox[t]{2mm}{\multirow{1}{*}{\rotatebox[origin=c]{90}{M:LSI-Co}}}
           & KNN        & 32-70(49) & 25-69(47) & 25-69(47) & 75-81(79) \\ 
           & NB         & 28-71(44) & 23-67(40) & 18-68(38) & 74-82(79) \\ 
           & SVM        & 37-71(51) & 21-67(41) & 22-67(41) & 74-87(81) \\ 
           & DT         & 22-75(46) & 22-75(46) & 21-75(46) & 63-75(70) \\ 
           & RF         & \textbf{39-77(59)} & \textbf{31-77(56)} & \textbf{32-77(56)} & \textbf{85-90(88)} \\ 
           & NN         & 33-71(49) & 29-66(45) & 29-66(45) & 77-88(83) \\ \midrule
\parbox[t]{2mm}{\multirow{1}{*}{\rotatebox[origin=c]{90}{M:BM25}}}
           & KNN        & 35-80(59) & 26-78(55) & 27-78(56) & 77-90(84) \\ 
           & NB         & 31-82(56) & 23-82(52) & 22-82(51) & 79-90(85) \\ 
           & SVM        & \textbf{55-86(74)} & 36-85(65) & 39-85(66) & \textbf{93-96(95)} \\ 
           & DT         & 28-77(55) & 28-76(55) & 27-76(54) & 68-81(75) \\ 
           & RF         & 42-85(67) & 36-84(64) & 36-84(64) & 90-95(93) \\ 
           & NN         & 47-87(72) & \textbf{45-87(70)} & \textbf{45-87(71)} & \textbf{93-96(95)} \\ \bottomrule

\end{tabular}
\caption{Performance of methods across six classifiers across all multiclass classification settings, without oversampling. M = Method, C = Classifier, P = Precision, R = Recall, F = F1 measure, A = AUC, unit is percentage}
\label{tab:rq1}
\end{table}

\begin{figure}
    \centering
    \includegraphics[width=\columnwidth]{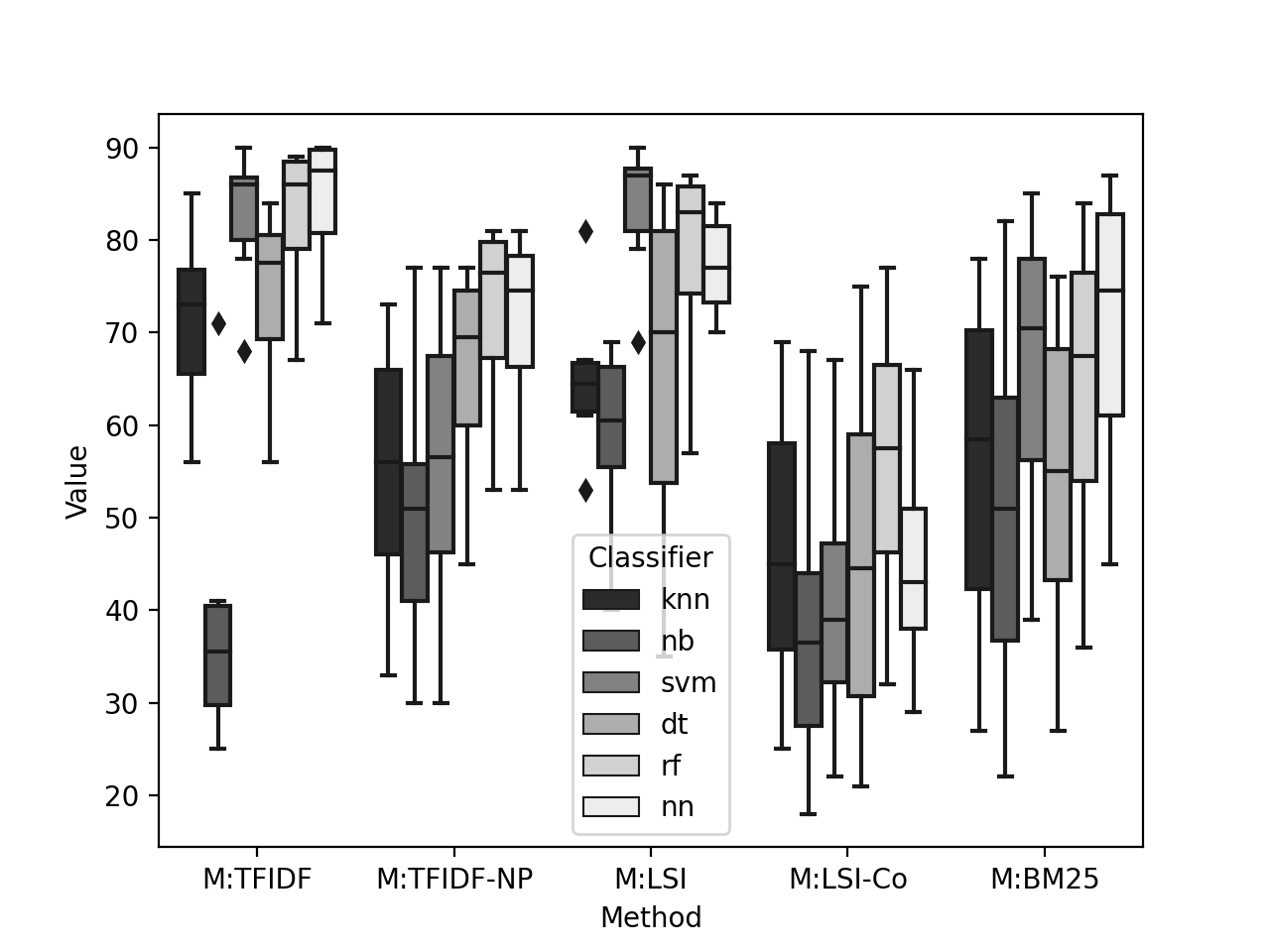}
    \caption{Boxplot of F1 score for five methods}
    \label{fig:rq1-f1-boxplot}
\end{figure}

We also report the boxplot of F1 score of for all implemented methods paired with each of the six classifiers in Figure~\ref{fig:rq1-f1-boxplot}. Each of the classifiers were run using all values of $n$. We discuss our observations from Table~\ref{tab:rq1} and Figure~\ref{fig:rq1-f1-boxplot}.

\textbf{SVM and NN classifiers work best for the M:TFIDF method.} We find SVM classifier shows the best performance in precision (87) and AUC score (99). The NN classifier shows the best performance in recall (84) and F1 score (84). SVM classifier differs by 3\% and 1\% in recall and F1 performance respectively compared with the NN classifier. The NN classifier differs by 1\% in both precision and AUC performance compared with the SVM classifier. The NB classifier performs the worst among the six classifiers. The F1 performance difference is around 40\% compared to NN and SVM classifiers. From Figure~\ref{fig:rq1-f1-boxplot}, we observe that SVM, RF, NN classifier shows \textit{close} Q1 (25th percentile) and Q3 (75th percentile) in F1 score. KNN and DT classifiers also shows \textit{close} Q1 and Q3 in F1 score, however, lagging behind the SVM, RF, and NN. NB has no overlap in Q1-Q3 range with rest of the five classifiers and performs the worst than rest of the classifiers.  

\textbf{RF classifiers work best for M:TFIDF-NP method.} We find RF classifier shows the best performance in all four metrics. NN classifier performs similar to RF classifier, however, differs by $1-2\%$ in four metrics compared to RF classifier. Moreover, NB classifier performs the worst among the six classifiers and the performance difference is around 20\% in F1 score compared to RF and NN classifiers. From Figure~\ref{fig:rq1-f1-boxplot}, we observe that all six classifiers have mutual overlap in Q1-Q3 range and the interquartile range is bigger than that of M:TFIDF.

\textbf{SVM classifiers work best for M:LSI method.} We find SVM classifier shows the best performance in all four metrics. RF and NN classifier performs similar to one another, however, differs by $5\%$ in f1 score compared to SVM classifier. Moreover, NB classifier performs the worst among the six classifiers in F1 score differing by 24\%. From Figure~\ref{fig:rq1-f1-boxplot}, we observe that RF and NN also follow SVM closely. Interquartile range varies among the six classifiers (i.e., DT and rest of the classifiers). 

\begin{figure}
    \centering
    \includegraphics[width=\columnwidth]{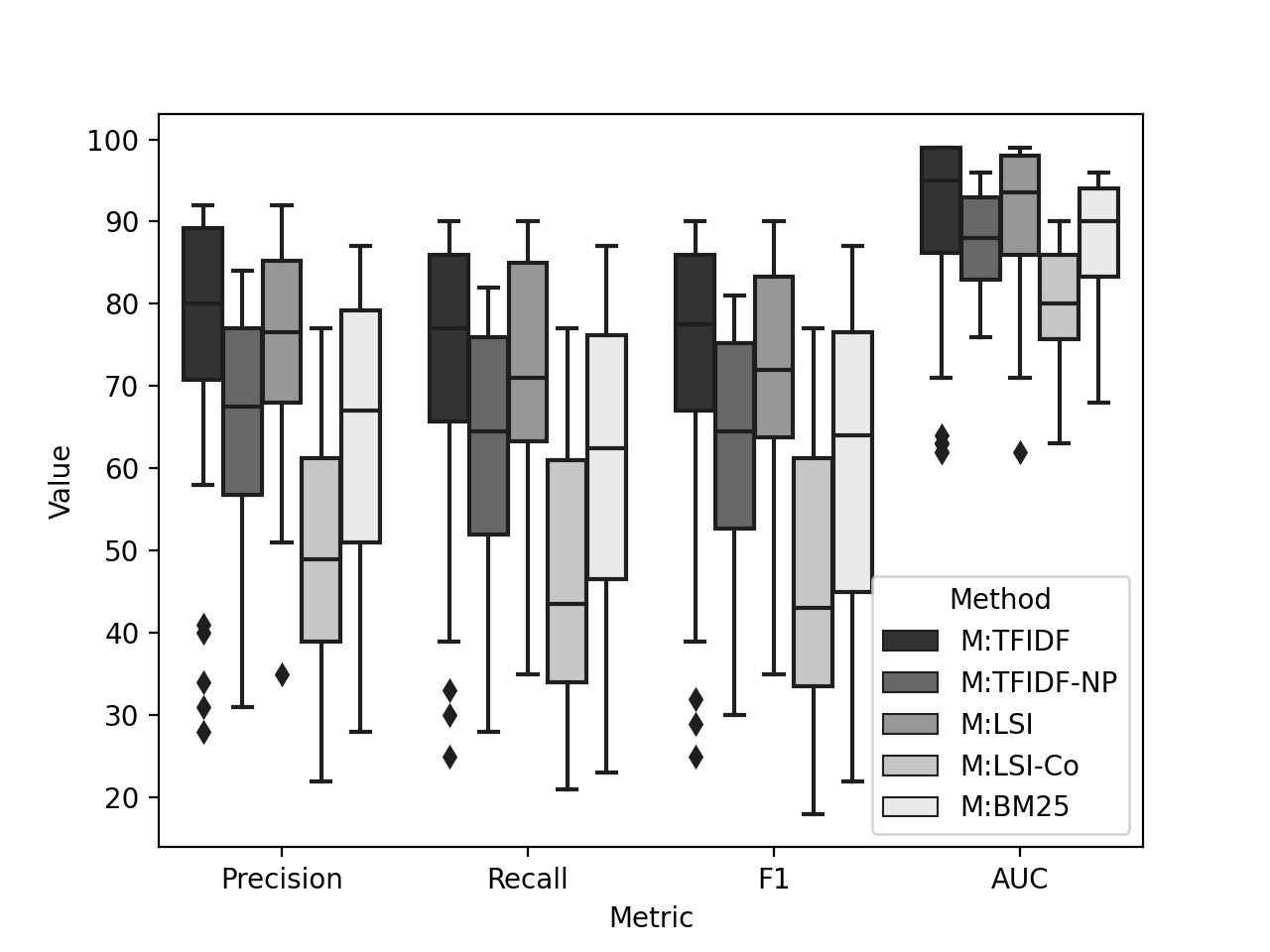}
    \caption{Boxplot of five methods on performance metrics}
    \label{fig:rq1-boxplot}
\end{figure}

\textbf{RF classifiers work best for M:LSI-Co method.} We find RF classifier shows the best performance in all four metrics. KNN and DT classifier performs similar to one another, however, differs by $10\%$ in f1 score compared to RF classifier. Moreover, NB classifier performs the worst among the six classifiers and the performance difference is around 18\% in F1 score compared to RF classifier. From Figure~\ref{fig:rq1-f1-boxplot}, we observe that RF is closely followed by KNN and DT. All six classifiers have mutual overlap with one another in Q1-Q3. 

\textbf{NN classifiers work best for M:BM25 method.} We find NN classifier shows the best performance in recall, F1 and AUC metrics. However, SVM classifier performs best in precision and equal to NN in AUC score. Moreover, NB classifier performs the worst among the six classifiers in F1 score differing by 20\%. From Figure~\ref{fig:rq1-f1-boxplot}, we observe that there are mutual overlaps among all six classifiers and the interquartile range is bigger than that of M:TFIDF and M:LSI.

\begin{table}[]
    \centering
    \footnotesize
    \begin{tabular}{llrrr}
    \toprule
        \multirow{2}{*}{\textbf{Method}} & \multirow{2}{*}{\textbf{Metric}} & \multicolumn{2}{c}{\textbf{Oversampling}} & \multirow{2}{*}{\textbf{Gain(\%)}}\\
        \cline{3-4}
        {} & {} & \textbf{No} & \textbf{Yes} & {} \\ \midrule
        \multirow{2}{*}{M:TFIDF} & F1 & 72 & 92 & 28 \\
         & AUC & 89 & 97 & 9 \\ \midrule
        \multirow{2}{*}{M:TFIDF-NP} & F1 & 62 & 82 & 32 \\
         & AUC & 87 & 95 & 9 \\ \midrule
         \multirow{2}{*}{M:LSI} & F1 & 71 & 89 & 20 \\
         & AUC & 91 & 96 & 5 \\ \midrule
        \multirow{2}{*}{M:LSI-Co} & F1 & 46 & 68 & 48 \\
         & AUC & 89 & 97 & 9 \\ \midrule
         \multirow{2}{*}{M:BM25} & F1 & 60 & 82 & 36 \\
         & AUC & 88 & 95 & 8 \\ 
     \bottomrule
    \end{tabular}
    \caption{F1 and AUC score of five methods with and without applying oversampling}
    \label{tab:rq2}
\end{table}




In Figure~\ref{fig:rq1-boxplot}, we report the boxplot of precision, recall, F1 and AUC score of the five methods run with all classifiers and classification settings. We list our observation below. 

\textbf{M:TFIDF and M:LSI performs best on TTP classification.} We observe that in all four metrics, M:TFIDF and M:LSI are ahead of the rest three methods. However, M:TFIDF shows slightly better performance than M:LSI. M:TFIDF-NP and M:BM25 show similar median score, however, M:BM25 shows a higher interquartile range than that of M:TFIDF-NP. M:LSI-Co lags behind the rest four methods. We also observe that the interquartile ranges of M:TFIDF and M:LSI are lower than the rest three methods. Finally, in AUC score, all methods are closer than the rest three metrics. 

\begin{tcolorbox}
The order of methods in terms of performance on highest performing classifier is: (i) M:TFIDF+(SVM or NN), (ii) M:LSI+SVM, (iii) M:BM25+NN, (iv) M:TFIDF-NP+RF, and (v) M:LSI-Co+RF
\end{tcolorbox}

\subsection{RQ2: Effect of class imbalance mitigation}
We report the average F1 and AUC scores of each method with and without oversampling run with six classifiers and six classification settings in Table~\ref{tab:rq2}. We list observations below. 

\begin{figure}
    \centering
    \includegraphics[width=\columnwidth]{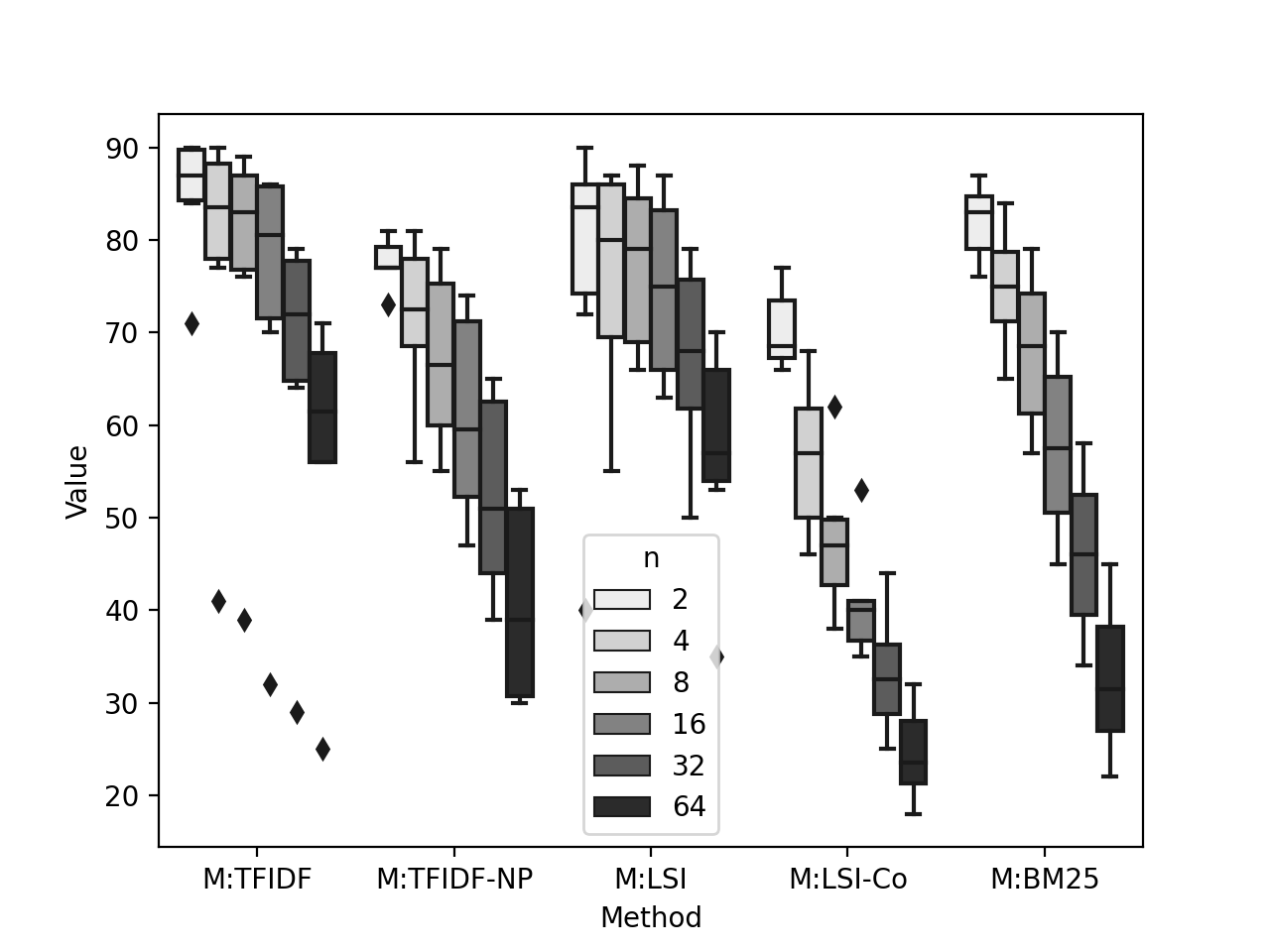}  
  \caption{F1 score of five methods, oversampling is not applied}
  \label{fig:rq3a}
\end{figure}

\textbf{Oversampling helps gaining more performance.} We observe that in the case of both F1 and AUC scores, all methods show better performance when oversampling is applied. In case of F1 score, the oversampling improved the score by 20 to 48\%. M:LSI-Co gained the most performance (48\%) and M:LSI gained the least (20\%). In case of AUC score, the oversampling improved the score by 5 to 9\%. M:LSI gained the least (5\%), and rest of the method gained 8-9\%. We also observe that the gain in F1 score is much higher than that of AUC score. 

\textbf{After applying oversampling, order of the methods in terms of performance remains similar.} Before applying oversampling, we see M:TFIDF shows the best F1 score, followed by M:LSI, M:TFIDF-NP, M:BM25, and M:LSI-CO. After applying oversampling, we observe the same order except M:TFIDF-NP and M:BM25 having a tie. Although M:BM25, M:LSI-Co, M:TFIDF-NP made the most gain in performance, these three methods still perform worse than M:TFIDF and M:LSI even after applying oversampling. 

\begin{tcolorbox}
The order of the methods in terms of F1 performance when oversampling is applied remains same before applying oversampling: (i) M:TFIDF, (ii) M:LSI, (iii) M:BM25, (iv) M:TFIDF-NP, and (v) M:LSI-Co. The performance gain in F1 score achieved in each method by oversampling is as follows in descending order: (i) M:LSI-Co, (ii) M:BM25, (iii) M:TFIDF-NP, (iv) M:LSI, and (v) M:TFIDF
\end{tcolorbox}

\subsection{RQ3: Effect of increase in class labels}
We report the boxplot of F1 and AUC scores using six classification settings ($n = 2, 4, 8, 16, 32, 64$) of all methods run with and without applying oversampling in Figure~\ref{fig:rq3a},~\ref{fig:rq3b},~\ref{fig:rq3c}, and~\ref{fig:rq3d}. We list our observations from the figure below.

\textbf{The F1 score monotonically decrease when oversampling was not applied.} We observe from Figure~\ref{fig:rq3a} that all the methods' F1 score drops strictly when $n$ increases, indicating that (i) methods perform better if the classifiers need to classify on two labels, (ii) methods show strictly decreasing order of performance when classifiers need to classify in multiclass classification, the more is the number of class labels, the less the performance is. We also observe that M:TFIDF and M:LSI show more robustness than the rest three methods when we increase the value of $n$. 

\begin{figure}
    \centering
    \includegraphics[width=\columnwidth]{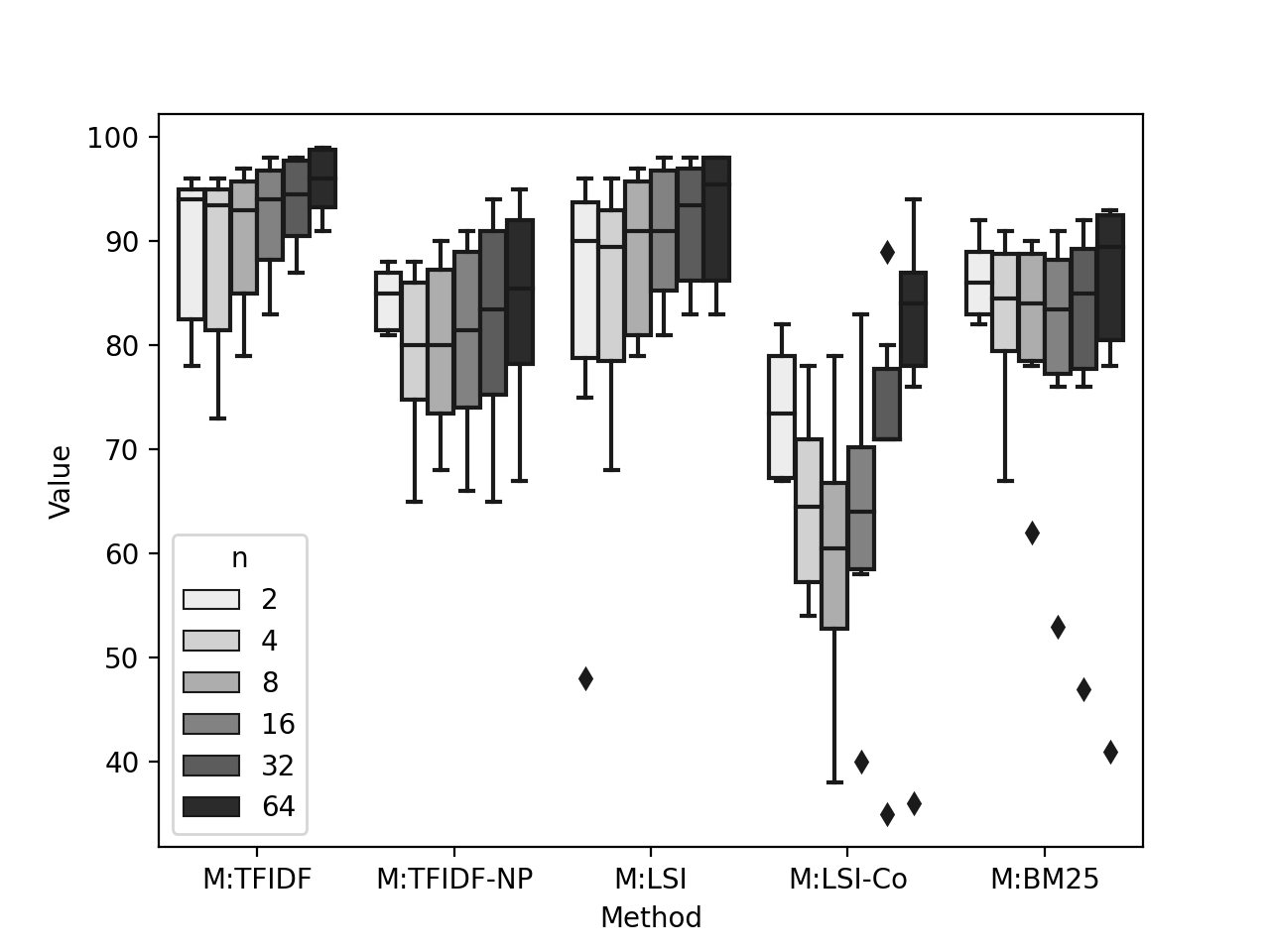}  
  \caption{F1 score of five methods, oversampling is applied}
  \label{fig:rq3b}
\end{figure}

\textbf{The F1 score does not monotonically increase or decrease when oversampling was applied.} We observe Figure~\ref{fig:rq3b} that, on oversmapled data, the methods behave differently to one another when we increase the value of $n$. M:TFIDF, M:TFIDF-NP, M:LSI, and M:BM25 show increase in performance, however, not in strict order. We also observe M:LSI-Co drops the performance from $n=2$ to $n=8$, however, then the performance increases bettering the case: $n=2$. 

\textbf{The AUC score first starts to increase and then decrease when oversampling was not applied}. We observe Figure~\ref{fig:rq3c} that, for all methods, with the increase of $n$, the scores first increase, reach a plateau, and then decrease. We observe that M:TFIDF shows the least variation of performance change, while M:LSI-Co shows the most variance. 

\textbf{The AUC score monotonically increase when oversampling was applied.} We observe Figure~\ref{fig:rq3d} that for all methods, AUC scores strictly increase with the increase of $n$. M:TFIDF-NP and M:LSI-Co made the most gain in AUC score. M:TFIDF demonstrates the least variance of AUC score with the increase of $n$.  

\begin{figure}
    \centering
    \includegraphics[width=\columnwidth]{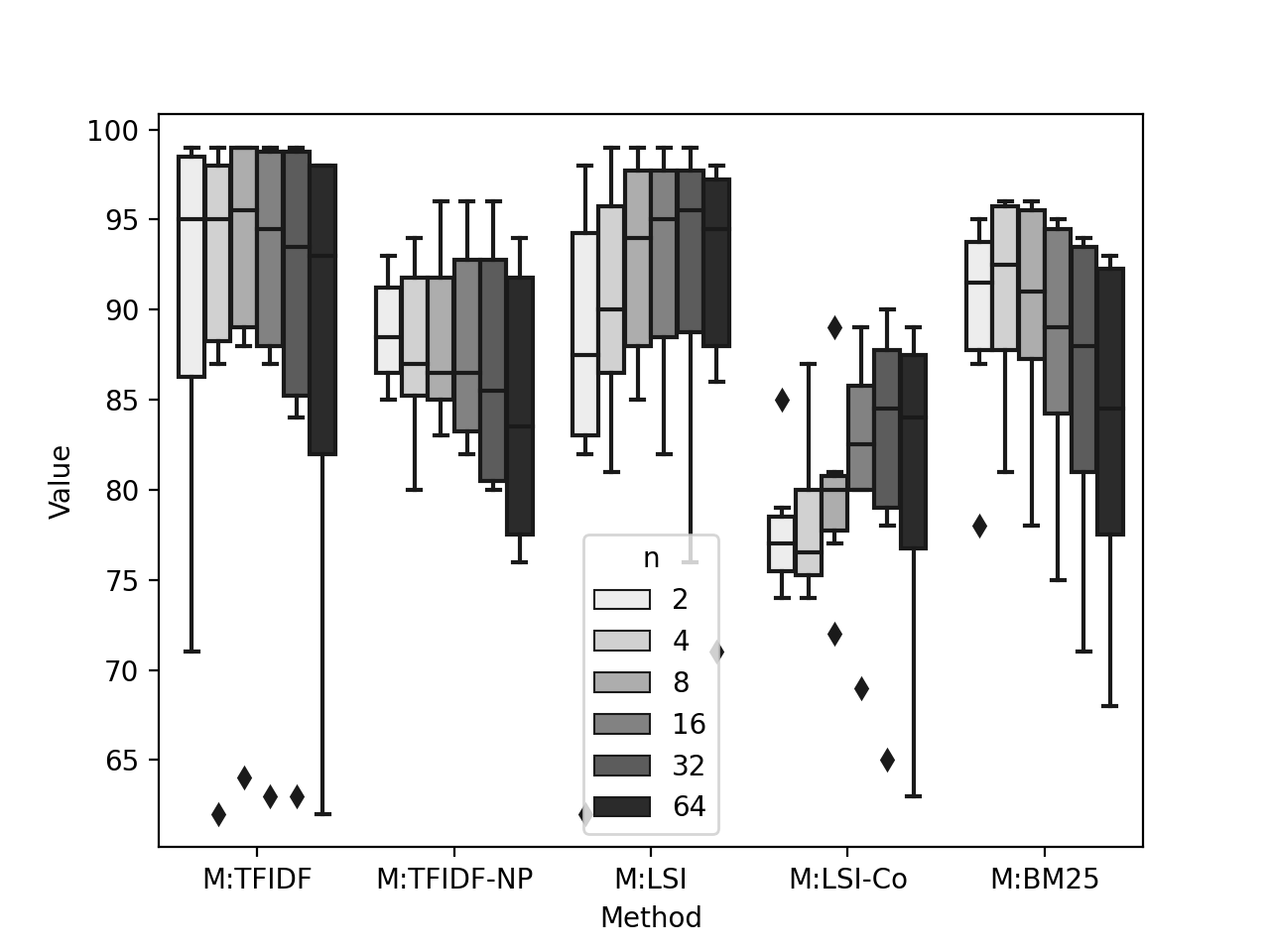}  
  \caption{AUC score of five methods, oversampling is not applied}
  \label{fig:rq3c}
\end{figure}

\begin{tcolorbox}
With the increase in $n$, F1 score decreases without oversampling. With oversampling, F1 score does not monotonically increase or decrease. With the increase in $n$, AUC score increases monotonically with oversampling, however, without oversampling, AUC score does not monotonically increase or decrease. 
\end{tcolorbox}

\section{Discussion} \label{discussion}
We observe the following order of performance for the compared methods: M:TFIDF, M:LSI, M:TFIDF-NP, M:BM25, M:LSI-Co. The performance rank of the methods does not change over different settings. We observe, with or without applying oversampling and across all multi-class classification settings, the M:TFIDF and M:LSI methods perform better than the other three methods. These two methods use the TFIDF vector from the whole corpus as features, while the other three methods use vectors of certain parts of speeches or similarity score. These results suggest that the specific existence of noun or verbs related to TTP is likely to be the most dominant feature for extracting TTP. 
    
Table~\ref{tab:reported-score} shows the maximum observed score from our implemented methods and the reported maximum performance score of the corresponding publications~\cite{noor2019machine, niakanlahiji2018natural, ayoade2018automated, li2019extraction, husari2017ttpdrill}. We observe similarity in the performance score of methods and corresponding studies. M:LSI and M:TFIDF shows 90\% F1 score while their corresponding work $S_1$, and $S_4$ shows F1 score of 96\% and accuracy score of 86\% respectively. M:LSI-Co is the least performing method from our observation and the reported score of the corresponding work $S_6$ is also associated with the least F1 score. In the case of M:BM25, we observe 87\% F1 score and the reported F1 score in the corresponding work $S_9$ is 86\%.  

\begin{figure}
    \centering
    \includegraphics[width=\columnwidth]{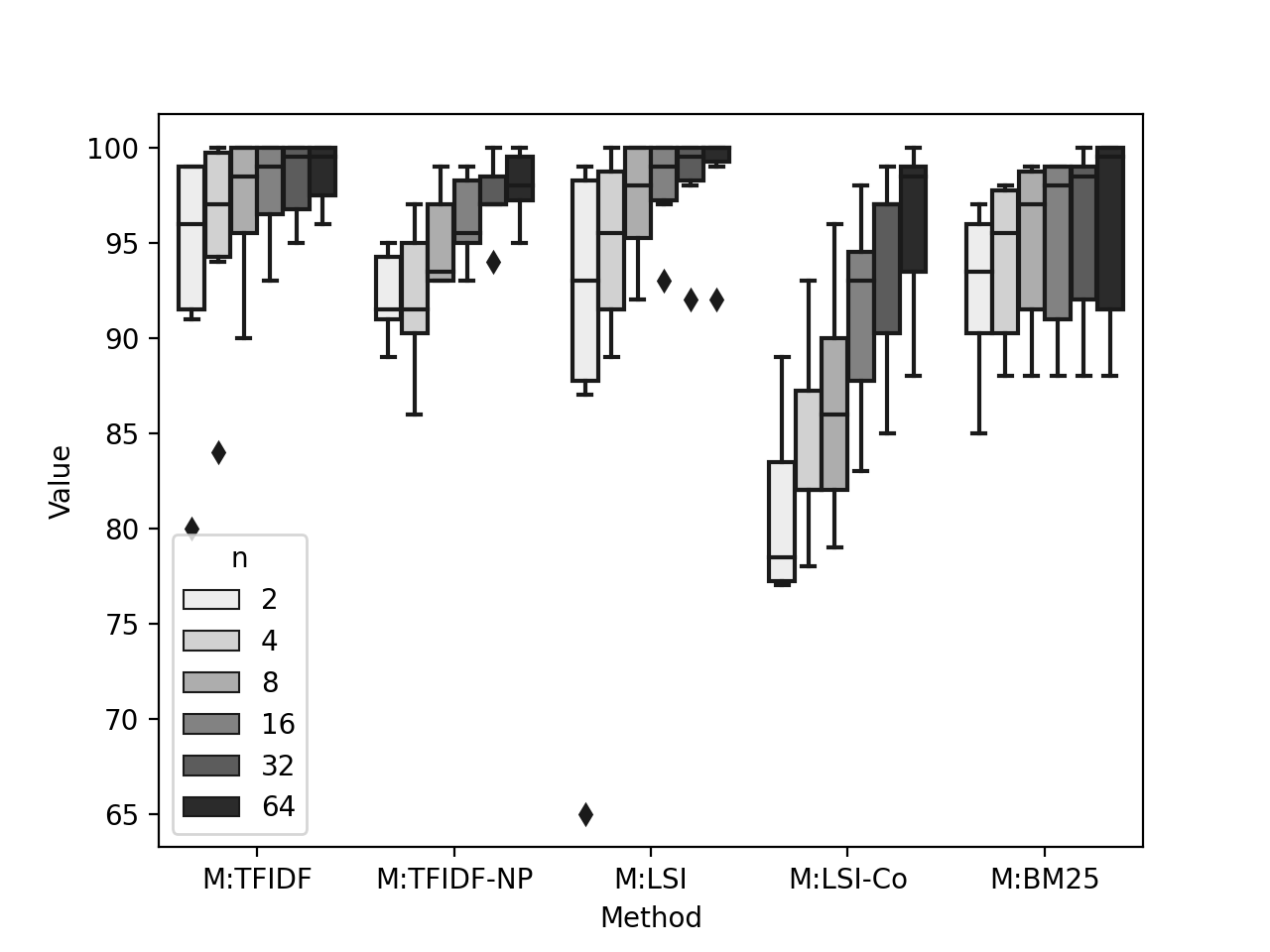}  
  \caption{AUC score of five methods, oversampling is applied}
  \label{fig:rq3d}
\end{figure}

Precision and recall performance do not vary by a large margin. We observe that precision and recall performance of each row in Table~\ref{tab:rq1} do not have more than 9\% difference, which is observed in the case of M:BM25 paired with SVM classifier. Our observations suggest that the methods offer a similar proportion of relevance and completeness while classifying TTP from a given text.
    
We observe that AUC scores of all methods are higher than F1 score of each method, indicating that (a) there could be a certain set of classification hyperparameters that would make the classification performance better; and (b) there could be bias in the dataset which impacts the precision and recall score.
    
    
We implement SVO tuples for comparing M:BM25 method, where we observe that our implementation for extracting SVO tuples is not perfect as the implementation cannot completely extract all words of objects in the case of: (i) when the object consists of multiple nouns; and (b) when the sentence is complex or written in passive voice. Improving the SVO tuple extraction, converting the passive voice to active voice, and breaking down a complex sentence to multiple simple sentences could make the performance better. 

\begin{table}[]
    \centering
    \footnotesize
    \begin{tabular}{lrlr}
    \toprule
        \textbf{Study} & \textbf{Reported Score(F1)} & \textbf{Implementation} & \textbf{Observed Score(F1)} \\ \midrule
        $S_1$ & 96 & M:LSI & 90 \\
        $S_2$ & - & M:TFIDF-NP & 81 \\
        $S_4$ & 86* & M:TFIDF & 90 \\
        $S_6$ & 57 & M:LSI-Co & 77 \\
        $S_9$ & 86 & M:BM25 & 87 \\
    \bottomrule
    \multicolumn{4}{l}{*: the reported score in $S_4$ is accuracy} \\
    \multicolumn{4}{l}{-: no performance score is reported in $S_2$} \\
    \end{tabular}
    \caption{Reported and observed performance score of compared methods}
    \label{tab:reported-score}
\end{table}

\section{Future research direction}
\label{future}
We advocate cybersecurity researchers and practitioners to investigate the performance of the discussed methods on applying a large corpus of threat reports. Moreover, we recommend establishing a benchmark dataset of large threat report corpus for conducting such experiments. Moreover, using NER, patterns of verb and noun co-occurrence, and regular expression of specific IoCs as features could have made the classification performance better. We recommend cybersecurity researchers investigate the performance benefits of incorporating these features while extracting TTP. As we discussed in Section~\ref{discussion}, existence of specific TTP related nouns and verbs could be the most dominant feature for TTP classification, and hence, we recommend researchers investigate the best performing textual features for the classification task through feature selection and ranking techniques. We run six classifiers with their default parameters, researchers can investigate the optimal set of hyperparameter~\cite{shu2021better} for each classifier for the best classification performance. Finally, classifying texts to corresponding tactics, procedures and applying multi-label classification (a sentence can be related to more than one TTP) could also be investigated. 

\section{Threats to validity}
\label{limitation}
We report the limitations we identify in this study. We did not compare the methods on corpus constructed from large threat reports. We also assume that one sentence is associated with only one corresponding attack technique. However, in threat reports, there could be sentences that might have more than one corresponding technique which can be classified with multi-label classifiers. For classification tasks, the dataset contains 64 technique names and hence, the classification performance for rest of the TTP is not evaluated. We also do not perform hyperparameter tuning for individual classifiers, which would have led to better performance. We also applied SMOTE technique to generate synthetic samples from numeric textual features, which might have introduced bias in the dataset. We did not perform a comparison study of all ten identified work and we also did not compare the five methods with tools from the industry such as~\cite{mitre-tram}. Finally, we implement the underlying methods from the study with our best effort, however there could be additional bias introduced along with our implementation. 



\section{Related work}
\label{related-work}
Rahman et al.~\cite{rahman2020literature} performed a systematic literature review on threat intelligence extraction from unstructured texts where they found 34 relevant studies and 8 data sources. The authors later surveyed 64 related studies in the extraction of threat intelligence from unstructured text where they identified ten CTI extraction goals~\cite{rahman2021attackers}. The authors proposed a generic pipeline for CTI extraction and surveyed the NLP and ML techniques utilized for extraction. Bridges et al.~\cite{bridges2017cybersecurity} performed a comparison study of prior work~\cite{bridges2013automatic, joshi2013extracting, jones2015towards} on cybersecurity entity extraction techniques. The authors used online blog articles on cybersecurity, National Vulnerability Database, and Common Vulnerabilities and Exposure (CVE) databases as corpus. The authors reported low recall of the compared methods and lack of published dataset. Our work differs from this work from Bridges et al. as our work compares the studies on TTP extraction while Bridges et al. compared the cybersecurity entity extraction from text. Tounsi et al.~\cite{tounsi2018survey} defined the four categories of CTI and discussed technical CTI, existing issues, emerging research, and trends in CTI domain. They also compared the features of existing CTI gathering and sharing tools. Wagner et al. ~\cite{wagner2019cyber} evaluated the technical and nontechnical challenges in state-of-the-art CTI sharing systems. Sauerwein et al.~\cite{sauerwein2017threat} studied 22 CTI-sharing platforms enabling automation of the generation, refinement, and examination of security data. Tuma et al.~\cite{tuma2018threat} conducted a systematic literature review on 26 methodologies on applicability, outcome, and ease of access of cyberthreat analysis. While these work focuses on CTI extraction and technical aspects of CTI tools, in our work, we focus on comparing TTP extraction methods to determine what method performs best and what are the areas for further enhancement.   

\section{Conclusion}
\label{conclusion}
In this work, we compare the underlying methods of five existing TTP extraction work~\cite{noor2019machine, niakanlahiji2018natural, ayoade2018automated, li2019extraction, husari2017ttpdrill} and the corresponding implementations are M:LSI, M:TFIDF-NP, M:TFIDF, M:LIS-Co, and M:BM25, respectively. We compared these methods on the performance of classifying textual descriptions of procedures to the corresponding technique names given in the ATT\&CK framework. From our experiment, we observe that: (a) M:TFIDF and M:LSI perform best in TTP classification from text; (b) performance of the methods drops when we increase the class labels in the classification task; and (c) oversampling improves the performance by mitigating the bias introduced by majority and minority classes in the dataset. We recommend (i) constructing an agreed-upon benchmark dataset; (ii) investigating all TTP extraction work from the literature and industry on large corpus dataset, and (iii) selecting optimal features for extracting TTPs from text. Cybersecurity researchers can use our work as a baseline for comparing and testing future TTP extraction methods.



\section*{Availability}
The dataset and source code of the implemented methods are available to download at this Github repository:~\cite{githubrepo}. 

\bibliographystyle{plain}
\bibliography{main}

\newpage \newpage

\section*{Appendix}
\label{appendix}
\begin{table*}[htb]
    \centering
    \footnotesize
    \begin{tabular}{lp{15cm}}
         \toprule
         \textbf{Id} & \textbf{Publication} \\
         
         $S_1$~\cite{noor2019machine} & Noor, Umara, Zahid Anwar, Tehmina Amjad, and Kim-Kwang Raymond Choo. "A machine learning-based FinTech cyberthreat attribution framework using high-level indicators of compromise." Future Generation Computer Systems 96 (2019): 227-242.  \\
         
         $S_2$~\cite{niakanlahiji2018natural} & Niakanlahiji, Amirreza, Jinpeng Wei, and Bei-Tseng Chu. "A natural language processing based trend analysis of advanced persistent threat techniques." In 2018 IEEE International Conference on Big Data (Big Data), pp. 2995-3000. IEEE, 2018.   \\
         
         $S_3$~\cite{ghazi2018supervised} & Ghazi, Yumna, Zahid Anwar, Rafia Mumtaz, Shahzad Saleem, and Ali Tahir. "A supervised machine learning based approach for automatically extracting high-level threat intelligence from unstructured sources." In 2018 International Conference on Frontiers of Information Technology (FIT), pp. 129-134. IEEE, 2018.   \\
         
         $S_4$~\cite{ayoade2018automated} & Ayoade, Gbadebo, Swarup Chandra, Latifur Khan, Kevin Hamlen, and Bhavani Thuraisingham. "Automated threat report classification over multi-source data." In 2018 IEEE 4th International Conference on Collaboration and Internet Computing (CIC), pp. 236-245. IEEE, 2018. \\
         
         $S_{5}$~\cite{piplai2020creating} & Piplai, Aritran, Sudip Mittal, Anupam Joshi, Tim Finin, James Holt, and Richard Zak. "Creating cybersecurity knowledge graphs from malware after action reports." IEEE Access 8 (2020): 211691-211703. \\
         
         $S_{6}$~\cite{li2019extraction} & Li, Mengming, Rongfeng Zheng, Liang Liu, and Pin Yang. "Extraction of Threat Actions from Threat-related Articles using Multi-Label Machine Learning Classification Method." In 2019 2nd International Conference on Safety Produce Informatization (IICSPI), pp. 428-431. IEEE, 2019. \\
        
        $S_{7}$~\cite{zhu2016featuresmith} & Zhu, Ziyun, and Tudor Dumitraş. "Featuresmith: Automatically engineering features for malware detection by mining the security literature." In Proceedings of the 2016 ACM SIGSAC Conference on Computer and Communications Security, pp. 767-778. 2016.  \\
        
        $S_{8}$~\cite{ramnani2017semi} 
        & Ramnani, Roshni R., Karthik Shivaram, and Shubhashis Sengupta. "Semi-automated information extraction from unstructured threat advisories." In Proceedings of the 10th Innovations in Software Engineering Conference, pp. 181-187. 2017. \\
        
        $S_{9}$~\cite{husari2017ttpdrill} 
        & Husari, Ghaith, Ehab Al-Shaer, Mohiuddin Ahmed, Bill Chu, and Xi Niu. "Ttpdrill: Automatic and accurate extraction of threat actions from unstructured text of cti sources." In Proceedings of the 33rd Annual Computer Security Applications Conference, pp. 103-115. 2017. \\
        
        $S_{10}$~\cite{husari2018using} 
        & Husari, Ghaith, Xi Niu, Bill Chu, and Ehab Al-Shaer. "Using entropy and mutual information to extract threat actions from cyber threat intelligence." In 2018 IEEE International Conference on Intelligence and Security Informatics (ISI), pp. 1-6. IEEE, 2018. \\
        
        \bottomrule
    \end{tabular}
    \caption{List of studies in the candidate set for the comparison}
    \label{tab:candidate-studies}
\end{table*}

\begin{table*}[]
\caption{Performance score of all the methods across six classifiers, six multiclass classification settings, and oversampling settings}
\label{tab:whole-report}
\centering
\footnotesize
\begin{tabular}{|c | c | c | cccc | cccc | cccc | cccc | cccc | cccc |}
\toprule
    \multirow{2}{*}{M} & \multirow{2}{*}{C} & \multirow{2}{*}{OS} & \multicolumn{4}{c}{n = 2} & \multicolumn{4}{c}{n = 4} & \multicolumn{4}{c}{n = 8} & \multicolumn{4}{c}{n = 16} & \multicolumn{4}{c}{n = 32} & \multicolumn{4}{c}{n = 64} \\ 
    \cline{4-7} \cline{8-11} \cline{12-15} \cline{16-19} \cline{20-23} \cline{24-27}
    {} & {} & {} & P & R & F & A & P & R & F & A & P & R & F & A & P & R & F & A & P & R & F & A & P & R & F & A \\ \midrule
    
    \multirow{12}{*}{M:TFIDF} & \multirow{2}{*}{KNN} & No & 88 & 84 & 85 & 93 & 80 & 76 & 77 & 92 & 79 & 76 & 76 & 92 & 76 & 69 & 70 & 91 & 70 & 63 & 64 & 89 & 63 & 55 & 56 & 88 \\
    
    {} & {} & Yes & 85 & 79 & 78 & 91 & 83 & 80 & 78 & 94 & 86 & 85 & 83 & 97 & 88 & 88 & 87 & 98 & 91 & 91 & 90 & 99 & 94 & 94 & 94 & 99 \\ \cline{2-27}
    
    {} & \multirow{2}{*}{NB} & No & 71 & 71 & 71 & 71 & 41 & 42 & 41 & 62 & 40 & 39 & 39 & 64 & 34 & 33 & 32 & 63 & 31 & 30 &	29 & 63 & 28 & 25 & 25 & 62 \\
    
    {} & {} & Yes & 81 & 79 & 79 & 80 & 74 &	74 &	73 &	84 & 80 &	80 &	79 &	90 & 85 &	85 &	83	& 93 &	88 &	88 &	87 &	95 &	92 &	92 &	91 &	97 \\ \cline{2-27}
    
    {} & \multirow{2}{*}{SVM} & No & 92 &	90 &	90 &	99 &	90 &	86 &	86 &	99 &	91 &	86 &	87 &	99 &	90 &	84 &	86 &	99 &	84 &	76 &	78 &	99 &	77 &	65 &	68 &	98 \\
    
    {} & {} & Yes & 96 &	96 &	96 &	99 &	96 &	96 &	96 &	100 &	97 &	97 &	97 &	100 &	98 &	97 &	98 &	100 &	98 &	98 &	98 &	100 &	99 &	99 &	99 &	100 \\ \cline{2-27}
    
    {} & \multirow{2}{*}{DT} & No & 89 &	84 &	84 &	84 &	86 &	81 &	81 &	87 &	81 &	80 &	79 &	88 &	78 &	77 &	76 &	87 &	69 &	68 &	67 &	84 &	58 &	56 &	56 &	80 \\
    
    {} & {} & Yes & 93 &	93 &	93 &	93 &	93 &	92 &	92 &	95 &	91 &	91 &	91 &	95 &	92 &	92 &	92 &	96 &	92 &	92 &	92 &	96 &	93 &	93 &	93 &	96 \\ \cline{2-27}
    
    {} & \multirow{2}{*}{RF} & No & 92 &	89 &	89 &	99 &	90 &	89 &	89 &	98 &	88 &	88 &	87 &	99 &	87 &	85 &	85 &	98 &	80 &	77 &	77 &	98 &	72 &	66 &	67 &	98 \\
    
    {} & {} & Yes & 95 &	95 &	95 &	99 &	95 &	95 &	95 &	99 &	95 &	95 &	95 &	100 &	96 &	96 &	96 &	100 &	97 &	97 &	97 &	100 &	98 &	98 &	98 &	100 \\ \cline{2-27}
    
    {} & \multirow{2}{*}{NN} & No & 91 & 90 & 90 & 97 & 91 & 89 & 90 & 98 & 90 & 89 & 89 & 99 & 87 & 87 & 86 & 99 & 80 & 79 & 79 & 99 & 74 & 71 & 71 & 98 \\
    
    {} & {} & Yes & 95 & 95 & 95 & 99 & 95 & 95 & 95 & 100 & 97 & 96 & 96 & 100 & 98 & 98 & 97 & 100 & 98 & 98 & 98 & 100 & 99 & 99 & 99 & 100 \\ \cline{2-27} \cline{1-1}
    
    \multirow{12}{*}{M:TFIDF-NP} & \multirow{2}{*}{KNN} & No & 77 & 74 & 73 & 85 & 71 & 70 & 68 & 86 & 63 & 60 & 60 & 85 & 56 & 51 & 52 & 84 & 50 & 44 & 44 & 82 & 39 & 33 & 33 & 79 \\
    
    {} & {} & Yes & 82 & 81 & 81 & 89 & 75 & 74 & 74 & 91 & 76 & 75 & 75 & 94 & 77 & 77 & 77 & 95 & 80 & 80 & 79 & 97 & 82 & 82 & 82 & 97 \\ \cline{2-27}
    
    {} & \multirow{2}{*}{NB} & No & 83 & 76 & 77 & 88 & 60 & 58 & 56 & 80 & 58 & 57 & 55 & 83 & 49 & 52 & 47 & 82 & 43 & 42 & 39 & 80 & 31 & 35 & 30 & 76 \\
    
    {} & {} & Yes & 84 & 81 & 81 & 92 & 71 & 66 & 65 & 86 & 77 & 74 & 73 & 93 & 77 & 74 & 73 & 95 & 79 & 74 & 74 & 97 & 81 & 78 & 77 & 98 \\ \cline{2-27}
    
    {} & \multirow{2}{*}{SVM} & No & 80 & 78 & 77 & 89 & 71 & 70 & 70 & 88 & 66 & 60 & 60 & 88 & 62 & 52 & 53 & 89 & 55 & 42 & 44 & 89 & 42 & 28 & 30 & 88 \\
    
    {} & {} & Yes & 84 & 83 & 83 & 91 & 77 & 77 & 77 & 92 & 70 & 68 & 68 & 93 & 69 & 66 & 66 & 96 & 68 & 65 & 65 & 97 & 72 & 67 & 67 & 98 \\ \cline{2-27}
    
    {} & \multirow{2}{*}{DT} & No & 81 & 78 & 77 & 86 & 76 & 76 & 75 & 85 & 75 & 74 & 73 & 85 & 68 & 66 & 66 & 83 & 60 & 58 & 58 & 80 & 48 & 46 & 45 & 77 \\
    
    {} & {} & Yes & 88 & 87 & 87 & 91 & 84 & 84 & 83 & 90 & 86 & 85 & 85 & 93 & 87 & 86 & 86 & 93 & 88 & 88 & 88 & 94 & 89 & 89 & 89 & 95 \\ \cline{2-27}
    
    {} & \multirow{2}{*}{RF} & No & 84 & 81 & 80 & 93 & 82 & 82 & 81 & 94 & 81 & 80 & 79 & 96 & 77 & 74 & 74 & 96 & 69 & 65 & 65 & 96 & 58 & 53 & 53 & 94 \\
    
    {} & {} & Yes & 89 & 88 & 88 & 95 & 88 & 88 & 88 & 97 & 90 & 90 & 90 & 99 & 92 & 91 & 91 & 99 & 94 & 94 & 94 & 100 & 95 & 95 & 95 & 100 \\ \cline{2-27}
    
    {} & \multirow{2}{*}{NN} & No & 84 & 82 & 81 & 92 & 80 & 80 & 79 & 93 & 77 & 77 & 76 & 93 & 74 & 72 & 73 & 94 & 67 & 64 & 64 & 94 & 57 & 52 & 53 & 93 \\
    
    {} & {} & Yes & 88 & 87 & 87 & 95 & 87 & 87 & 87 & 96 & 89 & 88 & 88 & 98 & 90 & 90 & 90 & 99 & 92 & 92 & 92 & 99 & 93 & 93 & 93 & 100 \\ \cline{2-27} \cline{1-1}
    
    \multirow{12}{*}{M:LSI} & \multirow{2}{*}{KNN} & No & 84 & 81 & 81 & 89 & 71 & 67 & 67 & 86 & 70 & 66 & 66 & 87 & 70 & 61 & 63 & 87 & 67 & 59 & 61 & 87 & 61 & 51 & 53 & 86 \\
    
    {} & {} & Yes & 82 & 76 & 75 & 87 & 77 & 75 & 75 & 91 & 81 & 80 & 79 & 95 & 86 & 85 & 85 & 97 & 90 & 90 & 90 & 98 & 94 & 94 & 93 & 100 \\ \cline{2-27}
    
    {} & \multirow{2}{*}{NB} & No & 68 & 56 & 40 & 62 & 65 & 55 & 55 & 81 & 69 & 67 & 67 & 91 & 72 & 69 & 69 & 93 & 68 & 64 & 64 & 94 & 62 & 56 & 57 & 94 \\
    
    {} & {} & Yes & 75 & 58 & 48 & 65 & 74 & 68 & 68 & 89 & 82 & 79 & 79 & 96 & 84 & 80 & 81 & 98 & 86 & 81 & 83 & 99 & 86 & 82 & 83 & 99 \\ \cline{2-27}
    
    {} & \multirow{2}{*}{SVM} & No & 92 & 90 & 90 & 98 & 91 & 87 & 87 & 99 & 91 & 87 & 88 & 99 & 90 & 86 & 87 & 99 & 83 & 78 & 79 & 99 & 75 & 67 & 69 & 98 \\
    
    {} & {} & Yes & 96 & 96 & 96 & 99 & 96 & 96 & 96 & 100 & 97 & 97 & 97 & 100 & 98 & 98 & 98 & 100 & 98 & 98 & 98 & 100 & 98 & 98 & 98 & 100 \\ \cline{2-27}
    
    {} & \multirow{2}{*}{DT} & No & 87 & 86 & 86 & 86 & 83 & 83 & 83 & 88 & 76 & 75 & 75 & 85 & 66 & 66 & 65 & 82 & 51 & 51 & 50 & 76 & 35 & 35 & 35 & 71 \\
    
    {} & {} & Yes & 91 & 90 & 90 & 90 & 89 & 89 & 89 & 93 & 87 & 87 & 87 & 92 & 86 & 86 & 86 & 93 & 85 & 85 & 85 & 92 & 84 & 84 & 84 & 92 \\ \cline{2-27}
    
    {} & \multirow{2}{*}{RF} & No & 90 & 86 & 86 & 96 & 90 & 86 & 87 & 97 & 88 & 85 & 85 & 98 & 85 & 80 & 81 & 97 & 78 & 70 & 72 & 97 & 67 & 55 & 57 & 95 \\
    
    {} & {} & Yes & 95 & 95 & 95 & 99 & 94 & 94 & 94 & 99 & 95 & 95 & 95 & 100 & 96 & 96 & 96 & 100 & 97 & 97 & 97 & 100 & 98 & 98 & 98 & 100 \\ \cline{2-27}
    
    {} & \multirow{2}{*}{NN} & No & 77 & 72 & 72 & 82 & 78 & 77 & 77 & 92 & 85 & 85 & 83 & 97 & 86 & 85 & 84 & 98 & 79 & 77 & 77 & 98 & 72 & 70 & 70 & 98 \\
    
    {} & {} & Yes & 91 & 89 & 90 & 96 & 91 & 91 & 90 & 98 & 96 & 96 & 96 & 100 & 97 & 97 & 97 & 100 & 98 & 98 & 97 & 100 & 98 & 98 & 98 & 100 \\ \cline{2-27} \cline{1-1}
    
    \multirow{12}{*}{M:LSI-Co} & \multirow{2}{*}{KNN} & No & 70 & 69 & 69 & 79 & 61 & 61 & 61 & 81 & 50 & 50 & 49 & 80 & 45 & 41 & 41 & 80 & 38 & 35 & 34 & 78 & 32 & 25 & 25 & 75 \\
    
    {} & {} & Yes & 79 & 79 & 79 & 85 & 72 & 72 & 71 & 89 & 67 & 67 & 67 & 91 & 72 & 72 & 71 & 95 & 81 & 81 & 80 & 97 & 88 & 88 & 88 & 98 \\ \cline{2-27}
    
    {} & \multirow{2}{*}{NB} & No & 71 & 67 & 68 & 77 & 52 & 46 & 46 & 74 & 41 & 40 & 38 & 77 & 39 & 37 & 35 & 80 & 33 & 26 & 25 & 82 & 28 & 23 & 18 & 82 \\
    
    {} & {} & Yes & 71 & 68 & 67 & 77 & 56 & 54 & 54 & 78 & 41 & 41 & 38 & 79 & 44 & 40 & 40 & 86 & 45 & 34 & 35 & 88 & 47 & 37 & 36 & 92 \\ \cline{2-27}
    
    {} & \multirow{2}{*}{SVM} & No & 71 & 67 & 67 & 74 & 54 & 51 & 49 & 76 & 47 & 43 & 42 & 80 & 51 & 36 & 36 & 85 & 44 & 30 & 31 & 87 & 37 & 21 & 22 & 86 \\
    
    {} & {} & Yes & 71 & 69 & 68 & 77 & 61 & 59 & 58 & 82 & 57 & 55 & 55 & 87 & 62 & 57 & 58 & 93 & 74 & 71 & 71 & 97 & 85 & 83 & 84 & 99 \\ \cline{2-27}
    
    {} & \multirow{2}{*}{DT} & No & 75 & 75 & 75 & 75 & 62 & 63 & 62 & 75 & 50 & 50 & 50 & 72 & 40 & 39 & 39 & 69 & 28 & 28 & 28 & 65 & 22 & 22 & 21 & 63 \\
    
    {} & {} & Yes & 79 & 79 & 79 & 79 & 72 & 72 & 71 & 82 & 66 & 66 & 66 & 81 & 68 & 68 & 68 & 83 & 71 & 71 & 71 & 85 & 76 & 76 & 76 & 88 \\ \cline{2-27}
    
    {} & \multirow{2}{*}{RF} & No & 77 & 77 & 77 & 85 & 69 & 68 & 68 & 87 & 64 & 61 & 62 & 89 & 57 & 53 & 53 & 89 & 50 & 44 & 44 & 90 & 39 & 31 & 32 & 89 \\
    
    {} & {} & Yes & 82 & 82 & 82 & 89 & 78 & 78 & 78 & 93 & 80 & 80 & 79 & 96 & 83 & 83 & 83 & 98 & 89 & 89 & 89 & 99 & 94 & 94 & 94 & 100 \\ \cline{2-27}
    
    {} & \multirow{2}{*}{NN} & No & 71 & 66 & 66 & 77 & 56 & 54 & 53 & 77 & 48 & 45 & 45 & 81 & 48 & 41 & 41 & 86 & 40 & 37 & 37 & 88 & 33 & 29 & 29 & 88 \\
    
    {} & {} & Yes & 68 & 67 & 67 & 78 & 58 & 58 & 57 & 82 & 53 & 53 & 52 & 85 & 61 & 60 & 60 & 93 & 72 & 72 & 71 & 97 & 85 & 85 & 84 & 99 \\ \cline{2-27} \cline{1-1}
    
    \multirow{12}{*}{M:BM25} & \multirow{2}{*}{KNN} & No & 80 & 78 & 78 & 87 & 74 & 72 & 72 & 90 & 68 & 64 & 65 & 88 & 54 & 51 & 52 & 84 & 44 & 38 & 39 & 80 & 35 & 26 & 27 & 77 \\
    
    {} & {} & Yes & 83 & 82 & 82 & 91 & 80 & 79 & 79 & 94 & 80 & 80 & 80 & 96 & 82 & 82 & 81 & 97 & 84 & 84 & 83 & 98 & 89 & 89 & 88 & 99 \\ \cline{2-27}
    
    {} & \multirow{2}{*}{NB} & No & 82 & 82 & 82 & 90 & 67 & 68 & 65 & 87 & 62 & 58 & 57 & 87 & 51 & 47 & 45 & 85 & 41 & 35 & 34 & 84 & 31 & 23 & 22 & 79 \\
    
    {} & {} & Yes & 83 & 83 & 83 & 90 & 71 & 69 & 67 & 89 & 68 & 62 & 62 & 90 & 60 & 53 & 53 & 89 & 55 & 45 & 47 & 90 & 52 & 40 & 41 & 88 \\ \cline{2-27}
    
    {} & \multirow{2}{*}{SVM} & No & 86 & 85 & 85 & 94 & 84 & 79 & 79 & 96 & 79 & 74 & 75 & 96 & 74 & 63 & 66 & 95 & 67 & 51 & 53 & 94 & 55 & 36 & 39 & 93 \\
    
    {} & {} & Yes & 89 & 89 & 89 & 96 & 89 & 88 & 89 & 98 & 89 & 89 & 89 & 99 & 87 & 86 & 86 & 99 & 88 & 87 & 87 & 100 & 91 & 91 & 91 & 100 \\ \cline{2-27}
    
    {} & \multirow{2}{*}{DT} & No & 77 & 76 & 76 & 78 & 73 & 72 & 71 & 81 & 61 & 61 & 60 & 78 & 51 & 51 & 50 & 75 & 41 & 41 & 41 & 71 & 28 & 28 & 27 & 68 \\
    
    {} & {} & Yes & 84 & 83 & 83 & 85 & 81 & 80 & 81 & 88 & 78 & 78 & 78 & 88 & 76 & 76 & 76 & 88 & 76 & 76 & 76 & 88 & 78 & 78 & 78 & 89 \\ \cline{2-27}
    
    {} & \multirow{2}{*}{RF} & No & 85 & 84 & 84 & 93 & 81 & 77 & 78 & 95 & 74 & 72 & 72 & 94 & 65 & 62 & 63 & 93 & 55 & 50 & 51 & 92 & 42 & 36 & 36 & 90 \\
    
    {} & {} & Yes & 89 & 89 & 89 & 96 & 88 & 88 & 88 & 97 & 88 & 88 & 88 & 98 & 89 & 89 & 89 & 99 & 90 & 90 & 90 & 99 & 93 & 93 & 93 & 100 \\ \cline{2-27}
    
    {} & \multirow{2}{*}{NN} & No & 87 & 87 & 87 & 95 & 86 & 83 & 84 & 96 & 80 & 79 & 79 & 96 & 71 & 70 & 70 & 95 & 60 & 58 & 58 & 94 & 47 & 45 & 45 & 93 \\
    
    {} & {} & Yes & 92 & 92 & 92 & 97 & 92 & 91 & 91 & 98 & 91 & 90 & 90 & 99 & 92 & 91 & 91 & 99 & 92 & 92 & 92 & 99 & 93 & 93 & 93 & 100 \\
\bottomrule
\end{tabular}
\end{table*}

\end{document}